\documentclass[useAMs,a4paper]{mn2e}
\usepackage{savesym}
\usepackage{graphicx}
\expandafter\let\csname equation*\endcsname\relax
  \expandafter\let\csname endequation*\endcsname\relax 
\usepackage{subfig}
\usepackage{amsmath}
\usepackage{amssymb}
\usepackage{verbatim}
\usepackage[yyyymmdd,hhmmss]{datetime}
\usepackage{array}
\usepackage{times}
\usepackage[total={17.8cm,24.0cm},centering]{geometry} 
\usepackage{color}

\newcommand{\be}{\begin{equation}}
\newcommand{\beq}{\begin{equation}}
\newcommand{\ee}{\end{equation}}
\newcommand{\eeq}{\end{equation}}
\newcommand{\eea}{\end{eqnarray}}
\newcommand{\bea}{\begin{eqnarray}}

\title[Hard X-rays from TDEs with large BH masses]{Hard X-ray emission from a Compton scattering corona in large black hole mass tidal disruption events}
\author [Andrew Mummery, Steven A. Balbus]{Andrew Mummery\thanks{E-mail:
andrew.mummery@physics.ox.ac.uk}, {Steven A. Balbus}
\\
Oxford Astrophysics, Denys Wilkinson Building, Keble Road, Oxford, OX1 3RH, United Kingdom}
\begin{document}

\date{}

\pagerange{\pageref{firstpage}--\pageref{lastpage}} \pubyear{2021}

\maketitle

\label{firstpage}

\begin{abstract} 
We extend the relativistic time-dependent thin-disc TDE model to describe nonthermal ($2-10$ keV) X-ray emission produced by the Compton up-scattering of thermal disc photons by a compact electron corona, developing analytical and numerical models of the evolving nonthermal X-ray light curves. In the simplest cases, these {X-ray light curves} follow power-law profiles in time. We suggest that TDE discs act in many respects as scaled-up versions of XRB discs, and that such discs should undergo state transitions into harder accretion states.   XRB state transitions typically occur when the disc luminosity becomes roughly one percent of its Eddington value.  We show that if the same is true for TDE discs then this, in turn, implies that TDEs with nonthermal X-ray spectra should come preferentially from large-mass black holes.   The characteristic hard-state transition mass is $M_{\rm HS} \simeq 2\times10^7 M_\odot$.
Hence, subpopulations of thermal and nonthermal X-ray TDEs should come from systematically different black hole masses.  We demonstrate that the known populations of thermal and nonthermal X-ray TDEs do indeed come from different distributions of black hole masses. The null-hypothesis of identical black hole mass distributions is rejected {by  a two-sample Anderson-Darling test with a $p$-value $< 0.01$}.   Finally, we present a model for the X-ray rebrightening of TDEs at late times as they transition into the hard state. 
These models of evolving TDE light curves are the first to join both thermal and nonthermal X-ray components in a unified scenario.
\end{abstract}

\noindent


\section{Introduction}
The disruption of a star by the gravitational tidal force of a supermassive black hole, a so-called tidal disruption event (TDE), can produce bright flares from otherwise quiescent galactic nuclei.  {In the last two decades these flares have been observed across a wide range of observing frequencies, including hard X-rays (e.g. Cenko et al. 2012), soft X-rays (e.g. Greiner et al. 2000), optical and UV (e.g. Gezari et al. 2008, van Velzen et al. 2021), infrared (e.g. Jiang {\it et al}. 2016,  van Velzen et al. 2016b), and radio (e.g. Alexander et al. 2016). TDEs harbouring powerful radio and X-ray bright jets have also been discovered (e.g. Burrows et al. 2011).  } 

{This wealth of observational data has resulted in a variety of physical models being proposed for the evolution of TDE emission. 
While models of the lower energy emission (e.g. optical, radio) often invoke outflows launched in the earliest stages of the TDE (e.g. Metzger \& Stone 2016, Strubbe \& Quataert 2009), or, especially for radio emission, a jet {(van Velzen {\it et al}. 2016a)},  high energy emission (UV and X-ray), is much more likely to result from an optically thick accretion disc which forms from the disrupted stellar debris.   Slim accretion discs (those with finite aspect ratios) have been invoked to model the evolving X-ray spectra of some TDEs (Wen {\it et al}. 2020), and as an explanation of the widely varying observed properties of TDE light curves (Dai {\it et al}. 2018).   Not all X-ray bright TDEs are dominated by emission from an accretion disc, however;  the emission of the brightest X-ray sources almost certainly come from a relativistic jet (Burrows {\it et al}. 2011).   }

{Time-dependent thin accretion disc
modelling has been successful in reproducing} the evolving X-ray (Mummery \& Balbus 2020a), UV (van Velzen {\it et al}. 2019, Mummery \& Balbus 2020a), and optical (Mummery \& Balbus 2020b) light curves of many TDEs.    {All of these works, however, assume that the emission from the evolving accretion discs is purely thermal (``soft state'' emission, in the language of X-ray binaries).  }  While some TDEs (e.g., five recently discovered sources in van Velzen {\it et al}. 2021) have X-ray spectra which are well-described by quasi-thermal (blackbody superposition) spectra, others (e.g., XMMSL1 J0619, Saxton {\it et al}. 2014) have X-ray spectra which are well-modelled by nonthermal  ``hard-state'' emission, characterised by a power-law with energy dependence.  

{Sources of this spectral type have been modelled by applying a sequence of steady-state AGN models to each individual observation (e.g. Saxton {\it et al}. 2019),  adusting the fitting parameters of these steady-state models by allowing them to vary with each observation.  However, sequential steady-state modelling does not allow for the true self-consistent dynamical evolution of system parameters with time.   For example, in the sequential steady-state approach the mass accretion rate $\dot M$ can be chosen to take whichever value best fits each particular observation.   In reality, these disc parameters are physically determined by the initial conditions and the constraints of mass and angular momentum conservation.   }

In this work, we incorporate dynamical time evolution by generalising the relativistic time-dependent thin disc TDE model of Mummery \& Balbus (2020a) with the incorporation of nonthermal  X-ray emission.   The key physical assumption made is that this emission arises from Compton up-scattering of thermal disc photons by a compact electron corona in a region enveloping the innermost stable circular orbit (ISCO). {The disc parameters are free to evolve self-consistently with time without prescriptive fitting, and both nonthermal and thermal X-ray spectral components are produced.}

This work represents the first extension of the relativistic TDE disc model to describe the transition from ``soft state'' thermal disc emission  to ``hard state'' nonthermal X-ray emission, and it leads to a number of interesting predictions.   For example, we show that simple scaling arguments lead to a minimum black hole mass {at which TDEs will form in the hard state (HS),}  of {order} $M_{\rm HS} \sim 2\times 10^7 M_\odot$, with only a weak dependence on the disc properties and luminosity at which the transition occurs (eq. \ref{MHS}).    For larger central mass TDEs, which form in the hard accretion state, we develop a phenomenological (but quantitive) model of the evolving nonthermal emission, assuming a compact corona covering the inner accretion disc.  We show that the large time evolution of nonthermal  X-ray light curves is given by the following simple asymptotic behaviour 
\beq\label{anlc}
L_X(t) \propto t^{-(2 + \Gamma)n/4},
\eeq
where $\Gamma$ is the X-ray spectrum photon index {(see \S\ref{compt_description} below),} and $n$ is the bolometric luminosity decay exponent, $L_{\rm bol}(t) \sim t^{-n}$.  We also demonstrate (numerically) that the compact-corona model produces observable levels of nonthermal X-ray emission from TDEs around large mass black holes for which thermal X-ray emission would be undetectable, and verify that the developed analytic modelling accurately describes the full numerical analysis.
Finally,  we develop a model where the electron corona `switches on' at large times for TDE discs formed at larger Eddington ratios $l_{\rm max} \equiv L_{\rm bol, max}/L_{\rm edd} \sim 0.1-1$, but then transition into a harder accretion state as matter is accreted and the disc luminosity decays with time.   

This paper represents the second part of a four paper TDE unification scheme (Mummery \& Balbus 2021a, Mummery 2021a, b). The layout of the paper is as follows. In \S \ref{massstates} we {discuss the current observational evidence for the existence of accretion state transitions in TDE disc systems, and} present a simple analysis which demonstrates that hard-state emission is expected from TDEs with masses larger than a well-defined characteristic scale $M_{\rm HS}$.  In \S \ref{compt_description} we describe our compact-corona model for the production and evolution of nonthermal  X-ray emission.  In \S \ref{results} we present both numerical and analytical results describing the observed properties of nonthermal X-ray {emission.}   In \S \ref{population},  we compare the properties of the current population of nonthermal  and thermal X-ray TDEs with the predictions of the  model.  In \S \ref{rebirghten}, we develop a model of a `switching-on' X-ray corona, before concluding in \S \ref{conc}.

\section{The spectral state of TDEs around large mass black holes}\label{massstates}

\subsection{Overview}

It is {well-established} that black hole X-ray binaries (hereafter XRBs) undergo transitions from the `high-soft' state, characterised by thermal X-ray spectra, to the `low-hard' state, characterised by nonthermal  (power-law) X-ray spectra, when the mass accretion rate in Eddington units is of order $1\%$,  $\dot m \equiv \dot M/ \dot M_{\rm edd} \sim 10^{-2}$ (e.g. Fender \& Belloni 2004, Maccarone  2003).  {Although} there is evidence that AGN discs undergo similar state transitions (Maccarone {\it et al}. 2003), the significantly extended viscous evolution timescale of typical AGN discs {generally prevents such transitions from being directly observed. } 
While the central black hole masses of TDE discs are similar to those of AGN, the discs are generally much more compact than their AGN counterparts.  This 
leads to a much shorter evolution timescale.   TDE discs may act as scaled-up analogues of XRB discs, and thus exhibit X-ray state transitions as well. 

TDEs have been observed {to form} in both the low-hard state (e.g, XMMSL2 J1446, Saxton {\it et al}. 2019) and the high-soft state (e.g, ASASSN-14li, Brown {\it et al}. 2017).   
Compelling (albeit circumstantial) evidence for state transitions in TDE discs has recently been presented by Jonker {\it et al}. (2020). { Jonker {\it et al}.\ observed 4 TDE sources at late times, all of which had been previously undetected at X-ray energies.}   { Nonthermal X-ray emission was observed from three of these four sources.  Although these sources were dim and detailed spectral information unavailable, one source was well-modelled by a hard power law $\Gamma \sim 2.5$, and another by a much softer power-law $\Gamma \sim 4$.   (The third  source was too dim for a reliable power-law index to be determined).  } {The authors suggest  
that the early-time X-ray emission of each TDE in the high-soft state was below detectable levels, but that at later times the discs underwent a state transition into the low-hard state.    Thereafter, detectable nonthermal X-ray emission emerged.}

{Furthermore, there is evidence that one TDE, AT2018fyk (Wevers {\it et al}. 2019, Wevers {\it et al}. 2021), has transitioned between these two states at early times in its evolution.} This X-ray emission from this source was initially dominated by its thermal component, but after a break in observations it was later found that the X-rays were dominated by a power-law component, some $\sim 75$ days after the ource was first detected.   When viewed in combination with the results of Jonker {\it et al}. (2020), it seems likely that AT2018fyk is the first example of an observed TDE undergoing an accretion state transition.  These results are supportive of the idea that TDE discs behave like `scaled up' XRB discs.  This TDE-XRB connection will be central to the modelling in this paper.   

Wevers (2020) presents evidence for an underlying trend between TDE spectral state and system parameters,  demonstrating a correlation between the central black hole mass of seven X-ray bright TDEs and the degree to which the X-ray spectra is dominated by nonthermal power-law components.  TDEs associated with most massive SMBHs were observed in the hard state, and it was suggested that this may be evidence for TDE discs around more massive black holes evolving more rapidly.

{We suggest that it is unlikely that TDE discs around more massive black holes evolve more rapidly than similar discs around smaller mass black holes.   In the first paper of this series (Mummery \& Balbus 2021a, hereafter Paper I) we showed that the viscous evolution timescale $t_v$ of an accretion disc of mass $M_d$ and constant $\alpha$-parameter scales as $t_v \propto M^{8/3} M_d^{-2/3} \alpha^{-4/3}$ (eq. 93, Paper I).  The evolutionary timescale of discs around more massive black holes are therefore expected to be in fact significantly longer than their less massive counterparts.}  In this paper, we {put forth} an alternative interpretation of Wevers (2020) findings:  TDEs around more massive black holes {are more likely to }{\it form} in the hard state, a result stemming from  the sensitive dependence of the bolometric luminosity upon the black hole mass. {This is not to say that TDE discs are unable to transition between accretion states.   Indeed, we assume that TDE discs do undergo such state transitions, as has likely already been observed.  The point we stress here is that, given the strong dependence on black hole mass of a TDE discs bolometric luminosity (eq. \ref{edrat}), an individual TDE system is statistically unlikely to form a disc with its critical Eddington ratio close to the transitional scale $\dot m \sim 0.01$, as opposed to either $\dot m \gg 0.01$, or $\dot m \ll 0.01$.     Therefore the majority (but not necessarily all) of TDEs should be observed to be evolving in either one of the two (thermal or nonthermal) accretion states.   A more detailed scaling analysis of this argument is presented below.  }

\subsection{Scaling analysis}\label{scalan}
In Paper I, we demonstrate that the peak temperature in a time-dependent relativistic accretion disc depends sensitively upon the black hole mass.  Expressed in terms of the disc mass $M_d$, $\alpha$-parameter $\alpha$, and black hole mass $M$, the temperature scales as
\beq\label{temp}
T_p \propto \alpha^{1/3} M_d^{5/12} M^{-7/6} .
\eeq
This implies an Eddington ratio $l$ which scales as  (Paper I):
\beq\label{edrat}
l \equiv {L_{\rm bol, peak} \over L_{\rm edd}} \propto {\alpha^{4/3} M_d^{5/3} \over M^{11/3}} .
\eeq
Note the strong black hole mass dependence of $l \propto M^{-11/3}$.   Consider now a disc with mass $M_d$, a prescribed $\alpha$-parameter, a central black hole mass of $M_{\rm edd}$ corresponding to a peak (Eddington) bolometric luminosity of $l = 1$.   With this definition for $M_{\rm edd}$, an otherwise identical disc with a central black hole of mass $M_{\rm HS}$ (``Hard State'')  forms at an Eddington ratio $l_{\rm HS}$, where 
\beq
M_{\rm HS} = M_{\rm edd}\, l_{\rm HS}^{-3/11},
\eeq
where $M_{\rm edd}$ now embodies the $\alpha$ and $M_d$ dependencies.   
Thus, if TDE discs with Eddington ratios $l_{\rm HS} \sim 10^{-2}$ form predominantly in the hard state and produce observable nonthermal X-ray emission, we would tend to find such nonthermal X-ray TDEs around more massive black holes (i.e., in excess of $M_{\rm edd}$), which is in accord with Wevers (2020) findings.  

Because of the strong mass dependence of the Eddington ratio, the change in black hole mass required for this hard-state 
transition is generally rather small:  e.g.,  $l_{\rm HS}^{-3/11} \simeq 3.5$ for $l_{\rm HS} = 0.01$.   The black hole mass at which the peak bolometric disc luminosity exactly equals the Eddington luminosity can be determined numerically;  for a Schwarzschild black hole with disc mass $M_d = 0.5 M_\odot$ and $\alpha = 0.1$, $M_{\rm edd} \simeq 5\times10^6M_\odot$ (Paper I).   The simple argument presented above would then predict nonthermal emission to be observable from an otherwise identical 
disc around a black hole with mass  $M > M_{\rm HS} \simeq 2 \times 10^7 M_\odot$.

This argument has interesting observational implications.  We demonstrated in Paper I that thermal X-ray emission is much weaker from more massive black holes, which host cooler discs (eq. \ref{temp}).    To leading order, the X-ray flux is given by (Paper I, eq. 70):  
\beq\label{suppres}
F_X \sim m^{-1/3} \exp\left(-m^{7/6}\right), 
\eeq 
where $m$ is a dimensionless mass ratio $M/M_\star$, where $M_\star$ is a characteristic mass with a typical value of $ \sim 10^6 M_\odot$ in applications of interest.  This Wien-tail suppression naturally leads to an upper limit for the observable mass associated with thermal X-ray emission, $M_{\rm lim} \sim 10^7 M_\odot$.  However, at $M \simeq 2 \times10^7 M_\odot$ the peak bolometric luminosity of the disc solutions becomes of order $\sim 1\%$ the Eddington luminosity, and it is at this scale that transitions to the hard state are expected.  
This simple analysis suggests that the populations of thermal X-ray TDEs and nonthermal  X-ray TDEs should be drawn from black hole populations with systematically different masses, being either distributed above (nonthermal) or below (thermal) a characteristic black hole mass of order $\sim 10^7 M_\odot$ (Figure \ref{example_Schwarz_nonthermal}).

\section{Comptonized disc emission}\label{compt_description}
\begin{figure}
  \includegraphics[width=.5\textwidth]{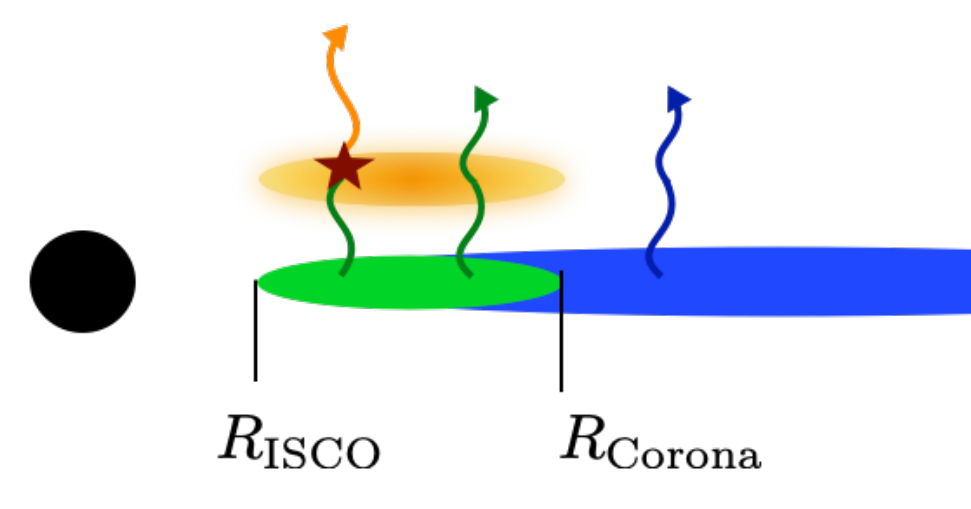} 
    \includegraphics[width=.5\textwidth]{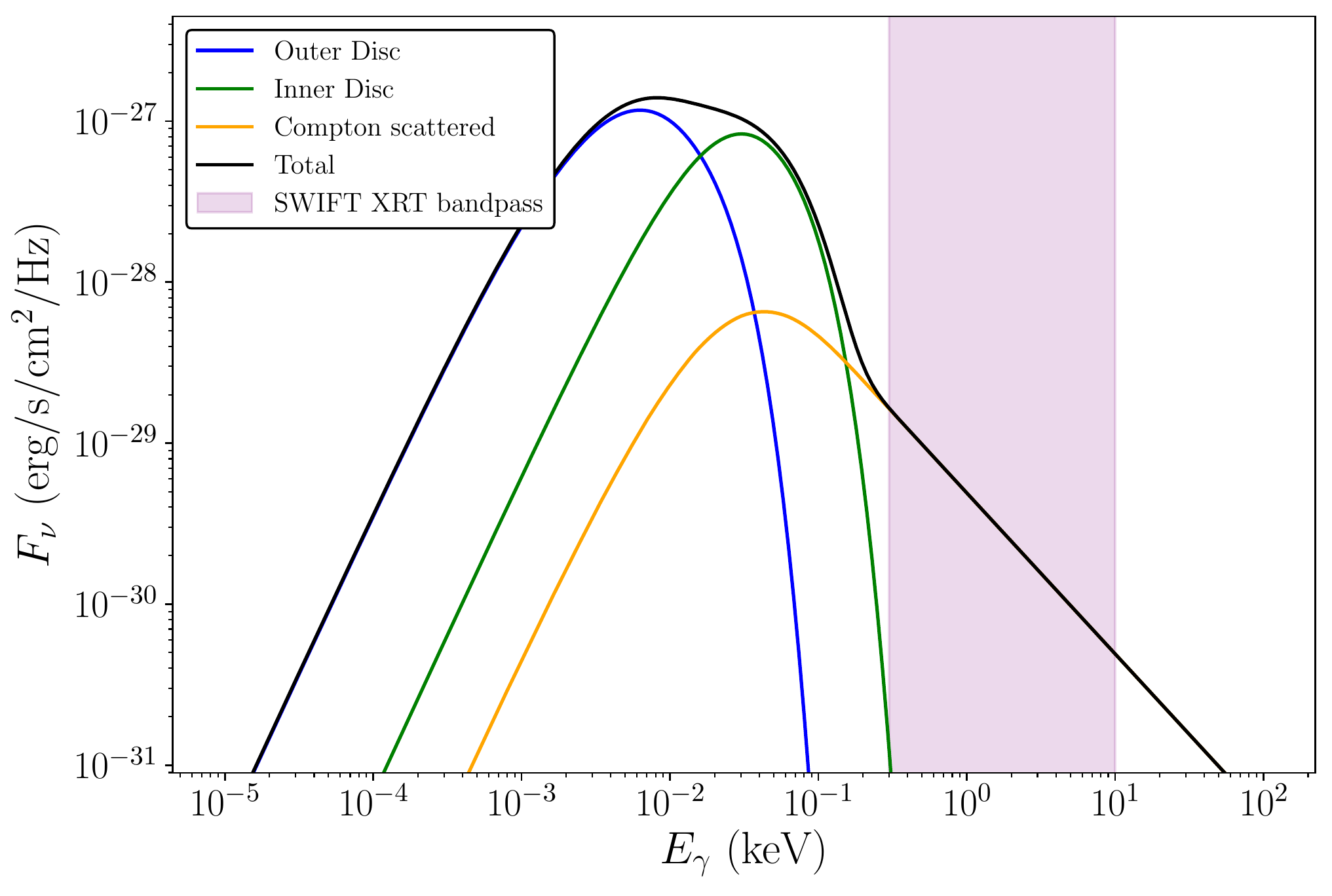} 
 \caption{Upper: A schematic of the Comptonizing corona geometry. Lower: an example `hard state' disc spectrum. This was  produced at the time when the discs bolometric luminosity peaks for a disc with initial mass $M_d = 0.5 M_\odot$ and $\alpha = 0.1$, around a Schwarzschild black hole with mass $M = 2\times10^7M_\odot$. The corona extended from the ISCO $(6r_g$) out to $R_{\rm Cor} = 12r_g$ and scattered a fraction $f_{SC} = 0.1$ of the soft disc photons with photon index $\Gamma = 2.0$. {The blue, green and orange curves in the lower panel correspond to the contributions to the total disc spectrum (black) from each of the different emission regions of the disc-corona system (see text). } } 
 \label{schematic}
\end{figure}

We model the effects of a Comptonizing corona following the SIMPL 
(Steiner {\it et al}. 2009) fitting module in XSPEC (Arnaud 1996).     We assume, without modelling in detail, that a Comptonising corona forms and is active when the Eddington ratio of the disc reaches some critical value, $l_{\rm HS}$, introduced in \S\ref{scalan}.    The corona covers the innermost regions  of the accretion disc, from the ISCO radius out to a radius $R_{\rm Cor}$.  The geometry is shown in the upper half of figure \ref{schematic}.   

Disc thermal photons emitted from radii $r > R_{\rm Cor}$ pass directly to the observer, as a standard colour-corrected blackbody superposition  (blue curve, fig. \ref{schematic}), a process discussed in Paper I, {with} photospheric corrections (Done {\it et al}. 2012).   For disc annuli $r \leq R_{\rm Cor}$ a fraction $1 - f_{SC}$ of the photons pass through the corona unscattered, contributing once again a colour-corrected black body component to the disc spectrum (green curve, fig. \ref{schematic}), while a fraction $f_{SC}$ undergo multiple Compton scattering (orange curve, fig. \ref{schematic}).  Photon number conservation in the scattering region then determines the observed properties of the scattered spectrum.  For an input distribution of photons $n(E_0){\rm d}E_0$, the outgoing photon distribution in an energy range $dE$ around energy $E$, $n'(E){\rm d}E$,  is given via an integral transformation  (Steiner {\it et al}. 2009)
\begin{multline}\label{scatter}
n'(E){\rm d}E = (1 - f_{SC}) n(E) {\rm d}E  \\  + f_{SC} \left[ \int_0^\infty n(E_0) G(E;E_0) {\rm d}E_0\right] {\rm d}E .
\end{multline}
In our model, the input distribution $n(E_0)$ is the photon occupation number at an energy $E_0$, obtained by summing the contribution from all of the disc annuli with $r < R_{\rm Cor}$, and assuming that each disc annulus emits thermally following the disc evolution solutions of Paper I.  The physics of the scattering events is then described by the Green's function $G(E;E_0)$, which accounts for the ensuing energy distribution of scattered photons for a $\delta$-function input at energy $E_0$.  
We use the following prescription for $G(E;E_0)$, which has a single free parameter $\Gamma$ (Sunyaev \& Titarchuk 1980; Steiner {\it et al}. 2009):
\beq\label{greens}
G(E;E_0) = {(\Gamma - 1)(\Gamma + 2) \over (1+2\Gamma)} {1 \over E_0}  
\begin{cases}
& \left({ E / E_0 }\right)^{1+\Gamma}, \quad E < E_0,  \\
\\
& \left({ E / E_0 }\right)^{-\Gamma}, \quad E \geq E_0 . 
\end{cases}
\eeq
The normalisation is determined by photon number conservation; the model is valid for any $\Gamma > 1$. 
This Green's function is a phenomenological 
model designed to recreate the properties of unsaturated repeated scattering of soft photons by non-relativistic thermal electrons.  It can be shown that the full solutions of the Kompaneets equation (which properly describes photon diffusion through the corona) do indeed produce power-law distributions of photon energies with this functional form (Shapiro, Lightman, \& Eardley 1976).  Technically, for any complete model of Comptonization, the up-scattered power-law distribution is cut off for photon energies larger than the temperature of the coronal electrons $E > k_BT_e$.  As we are considering relatively soft observed X-ray energies $E < 10$ keV, neglecting the cut-off in up-scattered photons above $k_B T_e$, which can be as high as $k_B T_e \sim 100{\rm' s}$ keV, should not significantly affect the results. 

A final -- and important -- assumption of our model is that the electron corona affects only
the observed distribution of photons via Compton scattering, and not the dynamical evolution of the accretion disc.  
Hard state models of XRBs often invoke truncations of the thin disc at large distances from the black hole (e.g. Narayan \& Yi 1995; Esin, McClintock \& Narayan 1997) 
with the central region replaced by a so-called advection-dominated accretion flow (ADAF).   AGN discs in quiescence ($L_{\rm bol} \sim 10^{-8} L_{\rm edd}$) are similarly expected to be truncated at very large radii from the central black hole.  On the other hand,  brighter AGN discs are often well-modelled by an accretion disc extending down to the ISCO, with innermost regions enshrounded by an electron scattering corona (e.g. the bright AGN Markarian 335, Wilkins \& Gallo 2015).    Measurements of the black hole spins of AGN (e.g., Reynolds 2013) are also consistent with the accretion disc extending down to the ISCO.  In this paper we shall assume that the underlying TDE accretion disc is described by the relativistic thin disc evolution equation (Balbus 2017) at all radii down to the ISCO.  No claims are made here that the model of the disc-corona geometry and evolution presented in this paper a {\it unique} description of a TDE disc in a hard accretion state, but it is one that is physically viable with testable predictions.  

\section{Results}\label{results}
In this section we present representative 
numerical and analytical results of the X-ray properties of hard-state disc dominated TDEs, assuming the disc-corona geometry of Figure \ref{schematic}.  We demonstrate that physically reasonable parameters for the disc and corona produce observable levels of nonthermal  X-ray flux for central black holes of large mass ($M > 2\times10^7M_\odot$).  We also present analytical results which describe the properties of the Comptonized X-ray flux well, both as a function of disc and black hole parameters, and as a function of time.  

The results in this section require the calculation of numerical solutions of the relativistic disc equation (Balbus 2017).   For this work we assume an $\alpha$-model for the turbulent stress, and the evolution equation is therefore non-linear. The disc is initially described by a ring of material, with total matter content $M_d$, at an initial radius $R_0 = 30 r_g$. This radius was chosen to roughly represent the characteristic circularisation radius of a TDE, which is anticipated to be $\sim 10$'s of gravitational radii. {(Note that the circularisation radius of a TDE is expected to be twice the incoming stars tidal radius.  We have found numerically that the precise value of $R_0$ had very little qualitative effect on the resultant disc light curves.)}   A full description of both the numerical problem to be solved, as well as the techniques used, may be found in \S4 of Paper I. 
 
\subsection{Observability}
\begin{figure}
    \includegraphics[width=.5\textwidth]{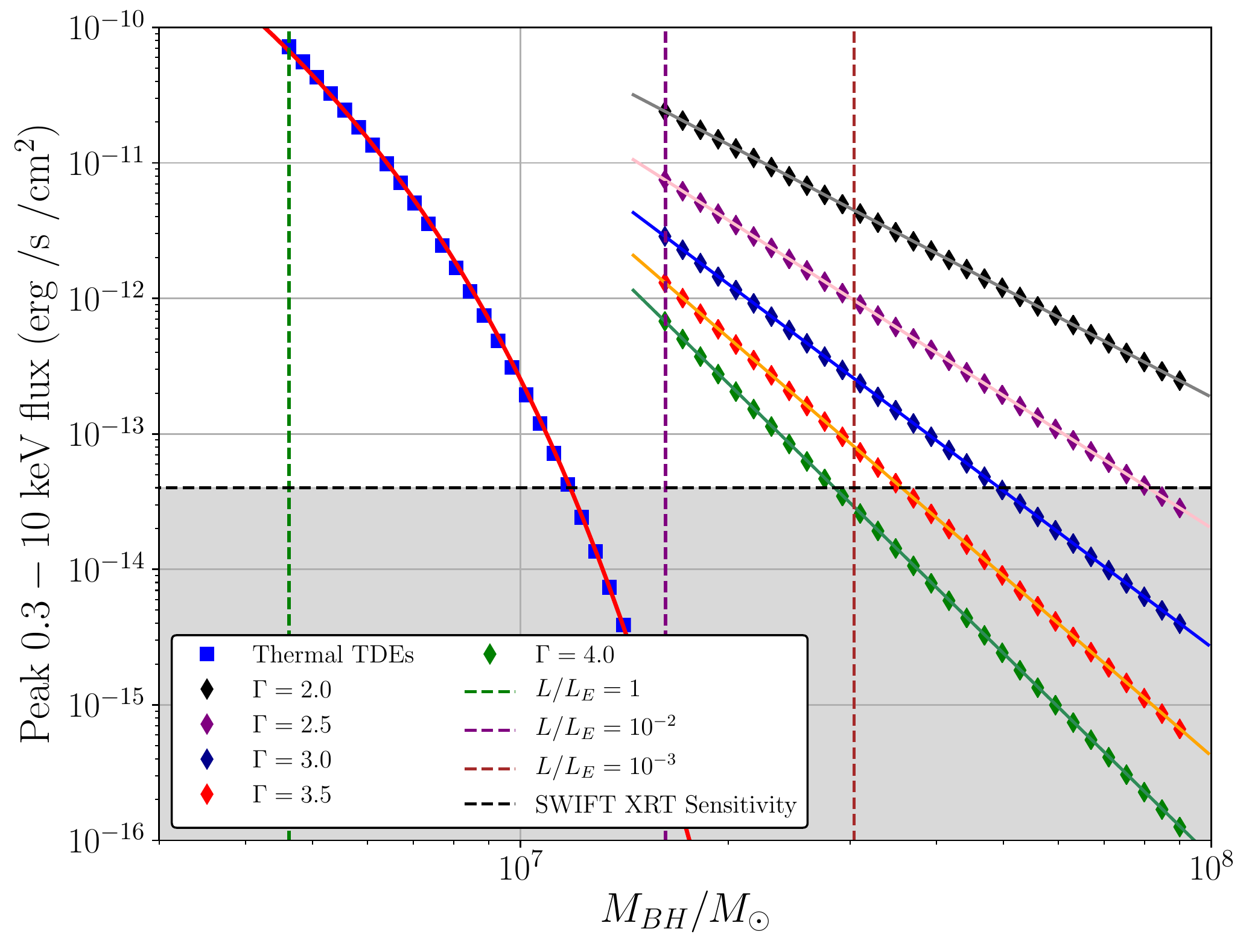} 
 \caption{The peak X-ray flux from evolving disc light curves with initial disc mass $M_d = 0.5 M_\odot$ and $\alpha = 0.1$ at a distance of $D = 100$ Mpc.  Thermal X-ray emission is strongly suppressed around larger mass black holes, becoming unobservable at $M \simeq 10^7 M_\odot$. This model produces observable nonthermal  X-ray fluxes from black holes of mass $M > 1.5\times10^7M_\odot$, where the bolometric Eddington ratio $l < 0.01$. We assumed that the corona extended from the ISCO $(6r_g$) out to $R_{\rm Cor} = 12r_g$ and scattered a fraction $f_{SC} = 0.3$ of the soft disc photons for a number of photon indices $\Gamma$, denoted on plot. The solid lines are analytical scaling relationships given by eq. \ref{scale}.The vertical dashed lines denote the Eddington ratio of the disc solutions. } 
 \label{example_Schwarz_nonthermal}
\end{figure}

Can a compact Compton-scattering corona  produce observable levels of X-ray flux from an accretion disc around a massive black hole which would otherwise be too cool to produce observable X-rays (eq. \ref{temp})?    
Figure \ref{example_Schwarz_nonthermal} shows the peak value of the evolving 0.3 -- 10 keV light curves of accretion discs, with {fixed} disc mass $M_d = 0.5 M_\odot$ and $\alpha = 0.1$, about Schwarzschild black holes of different masses, viewed from a distance $D = 100$ Mpc.  Points with blue squares are computed with pure thermal emission, as the Eddington ratio in these cases is $l > l_{\rm HS} = 10^{-2}$.   For black hole masses larger than $M \simeq 2 \times 10^7 M_\odot$, the disc Eddington ratio is $l < l_{\rm HS}$, and we include a Compton scattering corona covering the disc from the ISCO ($r_I = 6r_g$) out to $R_{\rm Cor} = 12 r_g$.  The corona scatters a fraction $f_{SC} = 0.3$ of the soft disc photons.  Light curves are computed for a number of different photon indices $\Gamma$. 

It is clear from Figure \ref{example_Schwarz_nonthermal}  that the compact corona model used here can readily produce 
observable levels of X-ray flux for TDE discs around large mass black holes.   The solid curves in the Compton scattering regime are the analytic scaling relationships derived in section \ref{analytic} (eq. \ref{scale}).   {Note that} coronae with harder indices (smaller $\Gamma$) produce {\em brighter} X-ray discs, observable at larger black hole masses.   {This model suggests} that the populations of thermal X-ray TDEs and nonthermal  X-ray TDEs should be drawn from black hole populations with systematically different masses, either above (nonthermal) or below (thermal) a characteristic black hole mass of order $\sim 10^7 M_\odot$.

\subsection{The observed X-ray spectral composition of TDE discs}
{The properties of the X-ray spectra of X-ray bright TDEs can be roughly split into three classes: those with `pure' thermal spectra (e.g. ASASSN-14li, Brown {\it et al}. 2017), those with pure nonthermal spectra (e.g.  XMMSL2 J1446, Saxton {\it et al}. 2019), and a third class observed with nonthermal spectral components accompanied by a low energy `thermal excess' (e.g. XMMSL1 J0740, Saxton {\it et al}. 2017). 

The results of our modelling thus far seem to imply that TDE discs should be observed in {\it either} a thermal or nonthermal  X-ray state (Figs \ref{schematic}, \ref{example_Schwarz_nonthermal}). In fact, there are at least five physical mechanisms by which a `thermal excess' may be observed in addition to a nonthermal power-law component in a TDE X-ray spectrum.  

The first mechanism lies beyond the scope of our simple model, and relates to the orientation angle of the TDE accretion disc with respect to the observers line of sight.  The material in the innermost regions of a black hole accretion disc is moving at highly relativistic speeds, and discs at moderate inclination angles therefore have regions where emitted photons undergo a large Doppler blueshift en route to the observer.  This blueshifted radiation will be observed at higher energies, which can result in an observed thermal excess. While this may seem like a minor effect, the thermal {\it Wien-portion of the X-ray flux } is exponentially sensitive to the redshift factor (defined by the ratio of the observed and emitted photon frequencies $f_\gamma = \nu_{\rm obs}/\nu_{\rm emit}$) of the hottest disc regions, which can vary by a factor $\sim 3$ with varying orientation angle. Modelling the spectral interaction between Doppler shifting, gravitational redshift, and Compton scattering of photons on their path to the observer is an extremely complex problem, which lies beyond the scope of the simple phenomenological model presented in section \ref{compt_description}.  

The remaining four mechanisms are more easily modelled and understood.  The first result arises from uncertainty in the magnitude of the colour-correction factor ($f_{\rm col}$) of the hottest disc regions. 
The vertical diffusion of disc photons can deviate from conditions of strict local thermal thermodynamic equilibrium (e.g.\ Done {\it et al}. 2012). 
Depending on the severity of this departure, this can result in the disc surface temperature deviating from the simple blackbody modelling assumptions of the underlying disc equations.   The scale of the deviation may be quantified by a simple correction factor,  $B_\nu(T) \rightarrow f_{\rm col}^{-4}B_\nu(f_{\rm col} T)$, where $B_\nu$ is the Planck function.   In producing figures \ref{schematic} and \ref{example_Schwarz_nonthermal}, we have taken a fairly standard value $f_{\rm col} = 2$;  however, there is uncertainty associated with the exact value.   Simple models (e.g. eq.\ [2] of Done {\it et al}.\ 2012) can produce values up to $f_{\rm col} \sim 2.7$.   A $\sim 30\%$ increase in the disc's colour-correction factor could lead to an observable thermal excess in some of the models discussed in this paper,  which nevertheless appear to be nonthermal in the X-ray band. 

TDE discs which transition into a harder accretion state at higher Eddington ratios $l_{\rm HS}$ can also show a prominent thermal excess.  While it is well-known that XRBs undergo transitions when the Eddington ratio is of order $1\%$, and there is some evidence that AGN discs undergo similar state transitions (Maccarone {\it et al}. 2003), the extended viscous evolution timescale of typical AGN discs makes it much more difficult to precisely determine a typical luminosity scale at which state transitions occur in discs around more massive black holes.   Initial findings specific to TDEs (Wevers 2020) suggest that these discs undergo a transition at $l_{\rm HS} \sim 0.03$, but with significant uncertainty.     If some TDE discs undergo state transitions at higher luminosity scales (e.g. $l_{\rm HS} \sim 0.1$), these higher-luminosity sources could well display a prominent thermal excess in addition to their coronal emission. This is a direct result of the exponential sensitivity of the thermal X-ray luminosity on the disc transitional luminosity scale (e.g. fig.\  [\ref{example_Schwarz_nonthermal}]). 

In addition, the physical properties of the corona itself (in particular its geometry and scattering fraction) strongly influence the nature of the thermal excess.    To see this most starkly, consider the (likely unphysical) limit $f_{SC} \rightarrow 1$.  The disc thermal emission would then appear to peak at the temperature $T_{\rm Cor} \equiv T(r = R_{\rm Cor})$, rather than $T_p = T(r=R_{\rm ISCO})$.  All photons emitted between $R_{\rm ISCO} < r < R_{\rm Cor}$ would Compton scatter.  Typically, the disc temperature falls off with radius like $T \sim r^{-3/4}$, meaning that $T_{\rm Cor}$ can be substantially lower than the peak disc temperature $T_{\rm Cor} \simeq T_p \, (R_{\rm ISCO} / R_{\rm Cor})^{3/4}$, particularly if the corona is large.   Conversely, lower scattering fractions and physically smaller corona will lead to a thermal disc profile much more similar to a `bare' disc, and are therefore more likely to display a thermal excess. 

}

\begin{figure}
    \includegraphics[width=.5\textwidth]{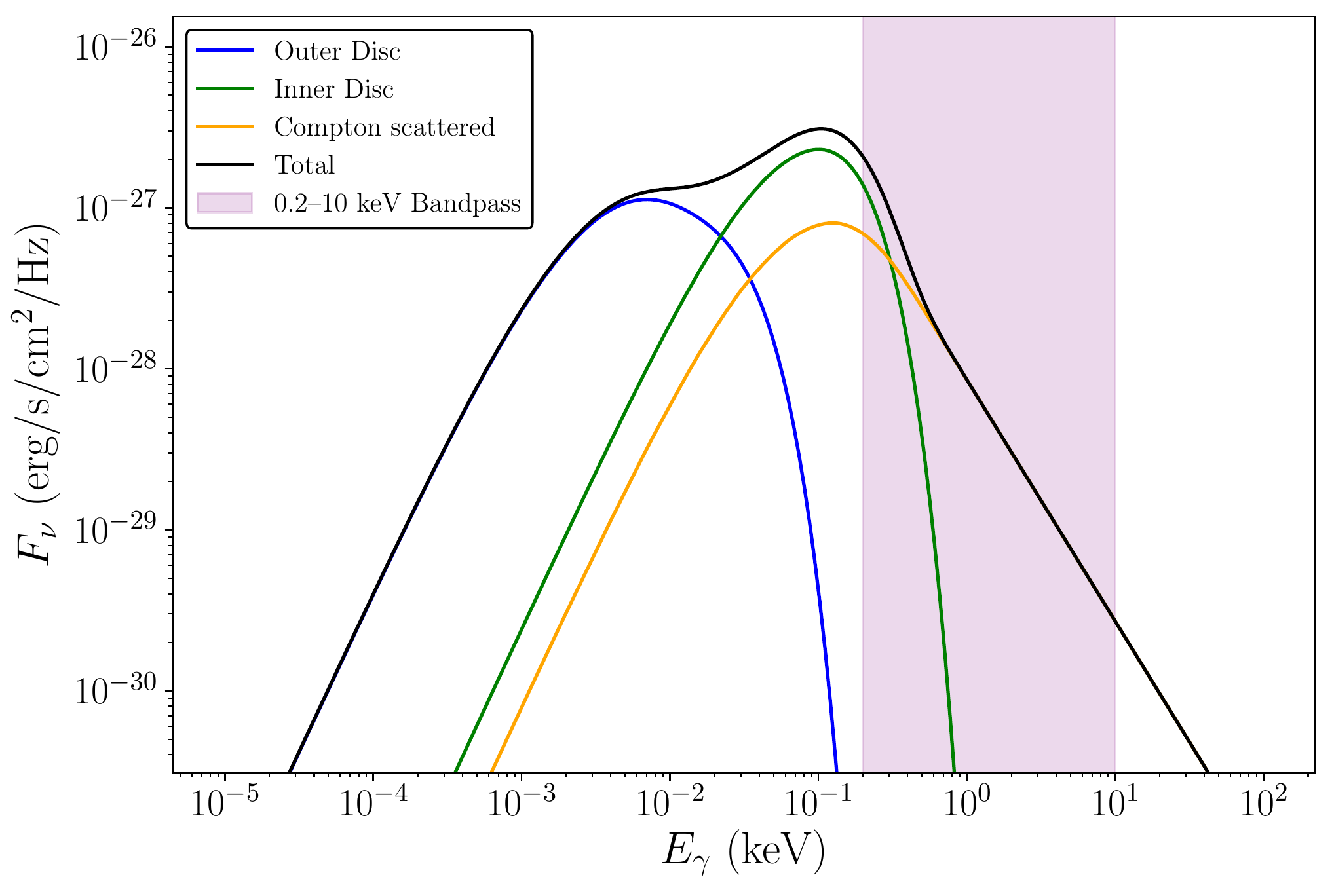} 
 \caption{This model produces an observable low energy `soft excess' above a dominant nonthermal  X-ray spectral component. This disc was described by mass $M_d = 0.5M_\odot$ and $\alpha = 0.1$, while the black hole had mass $M = 2\times10^7M_\odot$ and spin $a/r_g = 0.9$. The corona covered the disc from the ISCO to $R_{\rm Cor} = 8 r_g$, and scattered a fraction $f_{SC} = 0.3$ of the soft disc photons with $\Gamma = 2.5$.  The source-observer distance was $D = 100$ Mpc. A disc without a corona with these same parameters would have a bolometric luminosity $l = L_{\rm bol}/L_{\rm edd} = 0.03$. {The X-ray band displayed here ($0.2-10$ keV) corresponds to the XMM observing range; XMM has discovered a number of TDEs with mixed component X-ray spectra  (e.g. XMMSL1 J0740, Saxton {\it et al}. 2017). }} 
 \label{example_a09}
\end{figure}

{

Finally, and potentially most interestingly, black holes of higher spins will generally produce a more prominent thermal excess.   Under otherwise similar physical conditions, a large-spin Kerr black hole  emits more of its bolometric luminosity at higher energies, owing to the greater inward encroachment of the ISCO.  This leads to increased maximum disc temperatures.   Hence, discs around large-spin Kerr black holes with Eddington ratios $l \simeq 10^{-2}$ may still produce readily observable levels of thermal X-ray emission, in addition to their nonthermal power-law component.   

This is explicitly demonstrated in figure \ref{example_a09}.  If observed only in the X-ray band, the spectrum would appear to be well-modelled by a simple power-law with a  `soft-excess', which could be modelled as an additional black-body component with temperature $kT \sim 100$ eV.  Note that, other than a change in the spin, figure \ref{example_a09} retains the same black hole and disc parameters as seen in figure \ref{schematic}. The properties of this thermal excess can be further quantified by calculating its `temperature'.  This temperature measurement  is computed as follows.   The observed $0.3$--$10$ keV disc spectrum is modelled by two simple components, one thermal, the other nonthermal:
\beq
F(E) = A_0 E^{1-\Gamma} + A_1 B_\nu(E, T_{BB}), \quad 0.3 < E/{\rm keV} < 10,
\eeq
where $B_\nu(E, T_{BB})$ is a Planck function, defined with a single temperature $T_{BB}$, and $A_0$ and $A_1$ are constant amplitudes. By fitting the above functional form to a full disc spectrum, observed across the X-ray bandpass,  we obtain a best fitting temperature of the thermal excess $T_{BB}$.  Figure  \ref{TbbVSa} shows the best-fitting blackbody temperatures of a series of discs around black holes of increasing spins. Each disc had coronal parameters $f_{SC} = 0.3$, $\Gamma = 2.0$, $R_{\rm Cor} = 12 r_g$, a disc mass $M_d = 0.5M_\odot$ and $\alpha = 0.1$.    Figure \ref{TbbVSa} clearly shows that the inferred temperature of TDE accretion discs increase montonically with the central black holes spin. 
\begin{figure}
    \includegraphics[width=.5\textwidth]{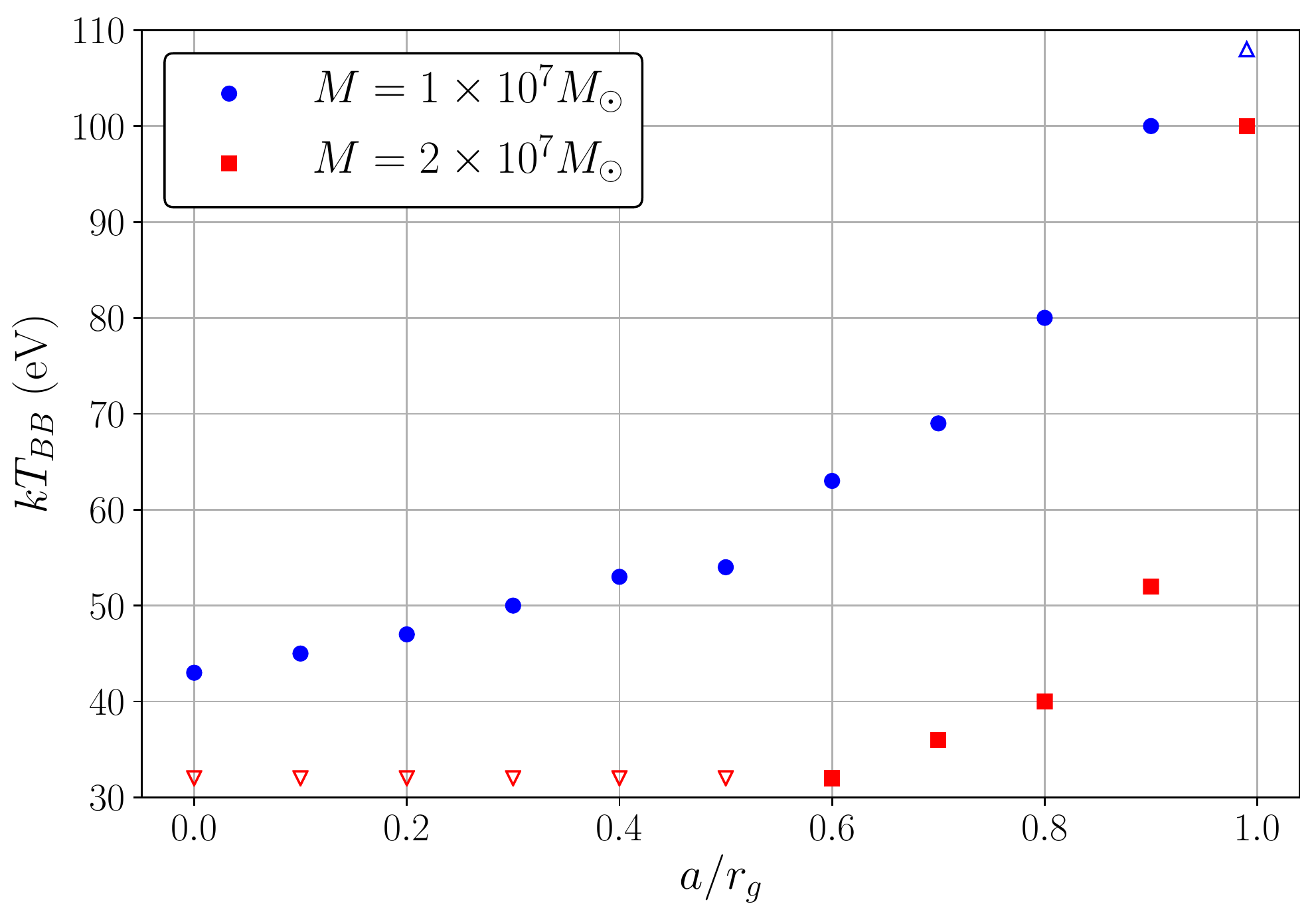} 
 \caption{{The `temperature' of the thermal excess of accretion discs (defined in text) around black holes of different spins of masses $M = 2\times 10^7 M_\odot$ (red squares), and $M=1\times 10^7M_\odot$ (blue circles). There is a clear positive trend between black hole spin and the inferred temperature of the thermal excess. For the highest mass black hole no thermal excess is present for spins $a/r_g < 0.6$. The temperature of the largest spin value ($a/r_g = 0.99$) for the lower mass black hole is not displayed for aesthetic reasons, but equals $kT_{BB} = 205$ eV.  }} 
 \label{TbbVSa}
\end{figure}

Clearly, there are many different mechanisms by which a hard-state TDE disc can produce a thermal excess.   It is likely that future high-quality observations of mixed thermal \& nonthermal X-ray TDEs (e.g.,  AT2018fyk [Wevers {\it et al}. 2019] and XMMSL1 J0740 [Saxton {\it et al}. 2017]) will be able to constrain, or even measure, the spins of supermassive black holes, a goal of considerable astrophysical interest. 

}

\subsection{X-ray flux scalings}\label{analytic}

Under the simplifying assumption that all of the observed X-ray flux comes from the hard Compton scattered component,  one may obtain expressions for this flux.  The observed flux across a band-pass with upper and lower energy limits $E_u$ and $E_l$ is 
\beq
F_X = {1\over D^2} \int_{E_l}^{E_u} E \, n'(E) \, {\rm d}E ,
\eeq
where $n'(E)$ is the photon occupation number as a function of energy, after any Compton scattering has taken place (eq. \ref{scatter}). 
The intrinsic disc spectrum will peak at some energy $E_p$ before falling off sharply.  If $E_l \gg E_p$, then only the Compton scattered photons will contribute to the observed X-ray flux (i.e., the second term of eq. \ref{scatter}), leading to  
\beq
F_X \simeq  {f_{\rm SC} \over D^2}   \int_{E_l}^{E_u} E  \left[ \int_0^\infty n(E_0) \,G(E;E_0) {\rm d}E_0\right] {\rm d}E .
\eeq
The photon occupation number from the intrinsic disc emission can be approximately described by the functional form 
\beq\label{discoc}
n(E_0) \simeq { N_\gamma \over E_p} \left({ E_0 \over E_p} \right)^b \exp\left(- { E_0 \over E_p }\right),
\eeq
where the cut-off energy is related to the peak disc temperature $T_p$ by $E_p \simeq k_B T_p$. This approximation for the intrinsic disc emission holds provided that the disc color-correction factor is a weak function of temperature.  
The normalisation $N_\gamma$ is (up to a numerical factor) the total number of photons emitted per unit time, and depends upn the disc and black hole properties.   Combining equations \ref{discoc} and \ref{greens}, we find 
\begin{multline}
F_X \simeq {1\over D^2} {(\Gamma - 1)(\Gamma + 2) \over (1+2\Gamma)} { N_\gamma f_{\rm SC} \over E_p} \int_{E_l}^{E_u} E^{1-\Gamma} {\rm d}E  \\
\int_0^\infty  \left({ E_0 \over E_p} \right)^b E_0^{\Gamma - 1} \, \exp\left(- { E_0 \over E_p }\right) {\rm d}E_0,
\end{multline}
which, after defining $ x \equiv E_0/E_p$, leaves
\begin{multline}\label{full_scaling}
F_X \simeq {1\over D^2} {(\Gamma - 1)(\Gamma + 2) \over (1+2\Gamma)} { N_\gamma f_{\rm SC}  E_p^{\Gamma - 1}} \int_{E_l}^{E_u} E^{1-\Gamma} {\rm d}E  \\
\int_0^\infty  x^{b + \Gamma - 1}\, \exp\left(- x\right) {\rm d}x .
\end{multline}
The integrals are mere normalisation factors, and the numerical value of the index $b$ does not affect the parameter dependence of the scattered X-ray flux.  With the total number of photons emitted per unit time $N_\gamma$ scaling as the bolometric disc luminosity divided by the characteristic disc energy,
\beq
N_\gamma \sim L_{\rm bol}/E_p \sim M^2 E_p^3,
\eeq
the parameter dependence of the X-ray flux is
\beq\label{scale}
F_X \propto f_{\rm SC} N_\gamma E_p^{\Gamma - 1} \propto   f_{\rm SC} M^2 E_p^{\Gamma + 2} .
\eeq
In Paper I we found that 
\beq
E_p \propto T_p \propto { M_d^{5/12} \alpha^{1/3} \over M^{7/6} } .
\eeq 
This gives the black hole mass dependence of the observed X-ray flux (cf.\ Fig. \ref{example_Schwarz_nonthermal}).   Recall that
after the disc temperature reaches its peak value it decays with a shallow power-law in time (e.g. Mummery \& Balbus 2020a) 
\beq
E_p \propto t^{-n/4},
\eeq
where $n$ is the bolometric luminosity decay index, $L \sim t^{-n}$.  This means that nonthermal  X-ray TDEs should be described by power-law decays at large times  
\beq\label{lc}
F_X(t) \propto t^{-(\Gamma+2)n/4} .
\eeq
The dependence of the X-ray decay index on $\Gamma$ means that softer (larger $\Gamma$) X-ray TDEs will decay more quickly than their harder counterparts, and for all corona with $\Gamma > 2$ the X-ray flux will evolve quicker than the bolometric disc luminosity. {It is interesting to note that this late-time decay law is much slower than the late-time decay of thermal X-ray TDE light curves: $F_X(t) \propto t^{-n/2} \exp\left(-At^{n/4}\right)$, where $A$ is a constant (Mummery \& Balbus 2020a). A general prediction of our model therefore is that thermal X-ray TDEs (or the thermal components of mixed spectrum TDEs)  will decay more rapidly than their nonthermal counterparts.  }

\begin{figure}
    \includegraphics[width=.5\textwidth]{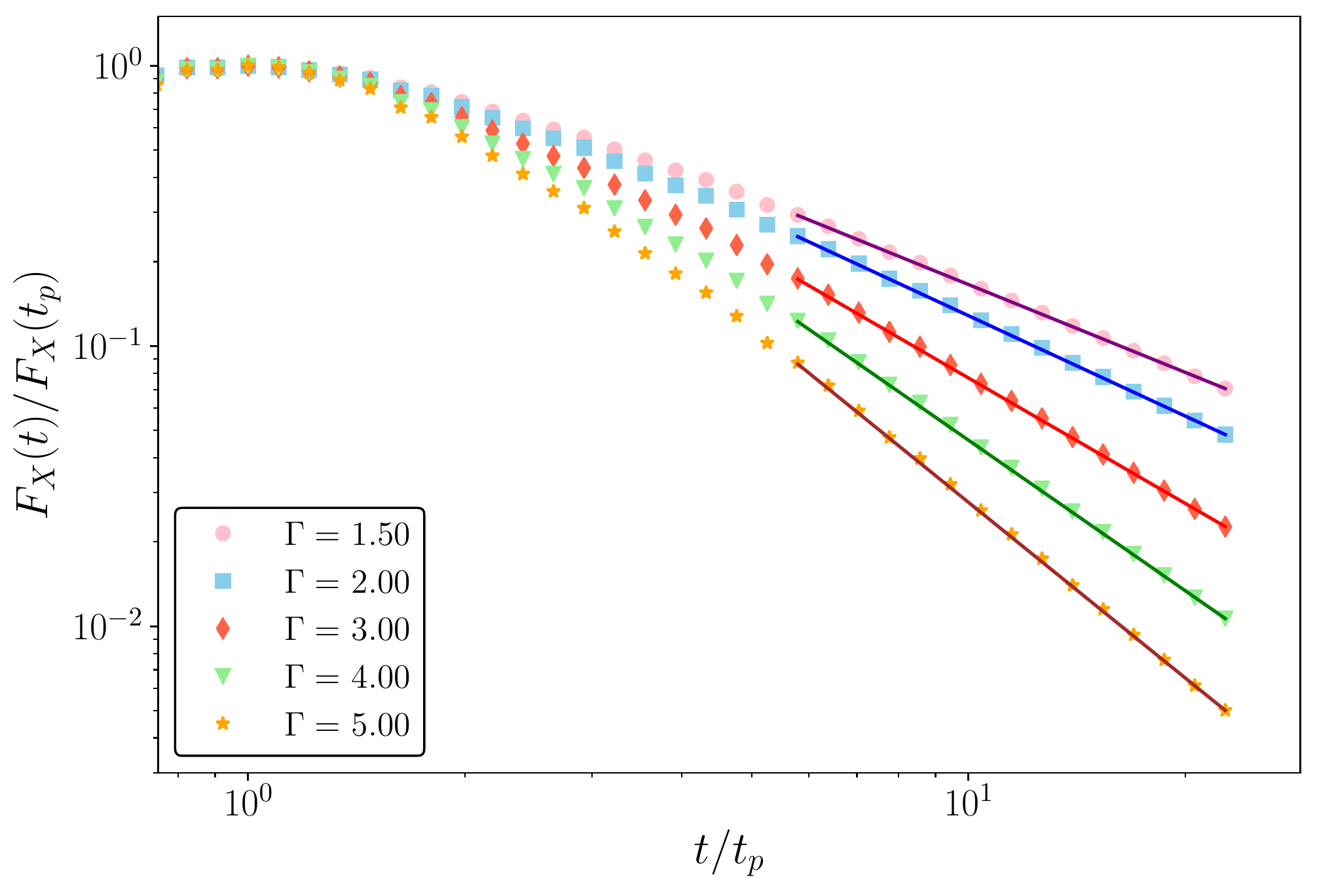} 
 \caption{Simple nonthermal  X-ray light curves from a black hole of mass $M = 2\times10^7M_\odot$, for discs with $M_d = 0.5 M_\odot$ and $\alpha = 0.1$, and a range of different corona photon indices $\Gamma$. The solid curves at large times show the predicted analytical behaviour (equation \ref{lc}) $F_X(t) \propto t^{-(2+\Gamma)n/4}$, with $n = 1.19$.   } 
 \label{fid_lc}
\end{figure}

Detailed numerical calculations of the time dependence of a hard component light curves agree very well with the simple analytic predictions of equation \ref{lc}.   In figure \ref{fid_lc} we compute the evolving 0.3 -- 10 keV flux for discs with initial mass $M_d = 0.5 M_\odot$ and $\alpha = 0.1$ around a Schwarzschild black hole of mass $M = 2 \times 10^7 M_\odot$. The corona extends from the ISCO ($6r_g$) to $R_{\rm Cor} = 12 r_g$, and scatters a fraction $f_{SC} = 0.3$ of the soft disc photons. Light curves are then produced for different photon indices $\Gamma$, using equation \ref{scatter}. For more details on the numerical solutions of the underlying relativistic disc equations see section 4 of Paper I. 

In Figure \ref{fid_lc} each light curve is normalized by its peak flux value and the time axis is plotted in units of the time at which the light curve peaked. The solid curves show the large time scaling result  $F_X(t) \propto t^{-(\Gamma + 2)n/4}$, with $n = 1.19$. This $n$ value is the theoretical value of the asymptotic decay index which is  appropriate at large times (Mummery \& Balbus 2019a, b). The analytical model describe the full numerical results at large times extremely well.  

{Finally, we note that two of the free parameters describing the disc-corona model, $R_{\rm Cor}$ and $f_{SC}$, exhibit a formal degeneracy:  a physically larger corona (larger $R_{\rm Cor}$) which scatters a smaller fraction $f_{SC}$ of the thermal disc photons can produce qualitatively similar X-ray fluxes to those  emerging from a smaller corona scattering a larger photon fraction.  However, in practice this is only a weak degeneracy, since the X-ray flux is rather insensitive to the physical size of the corona $R_{\rm Cor}$.   As can be seen from equation \ref{full_scaling}, the properties of the X-ray flux primarily depend on the coronal scattering fraction $f_{SC}$ and the power-law index $\Gamma$, a finding that has been numerically verified.  }

\subsection{Evolving coronal properties}
\begin{figure}
    \includegraphics[width=.5\textwidth]{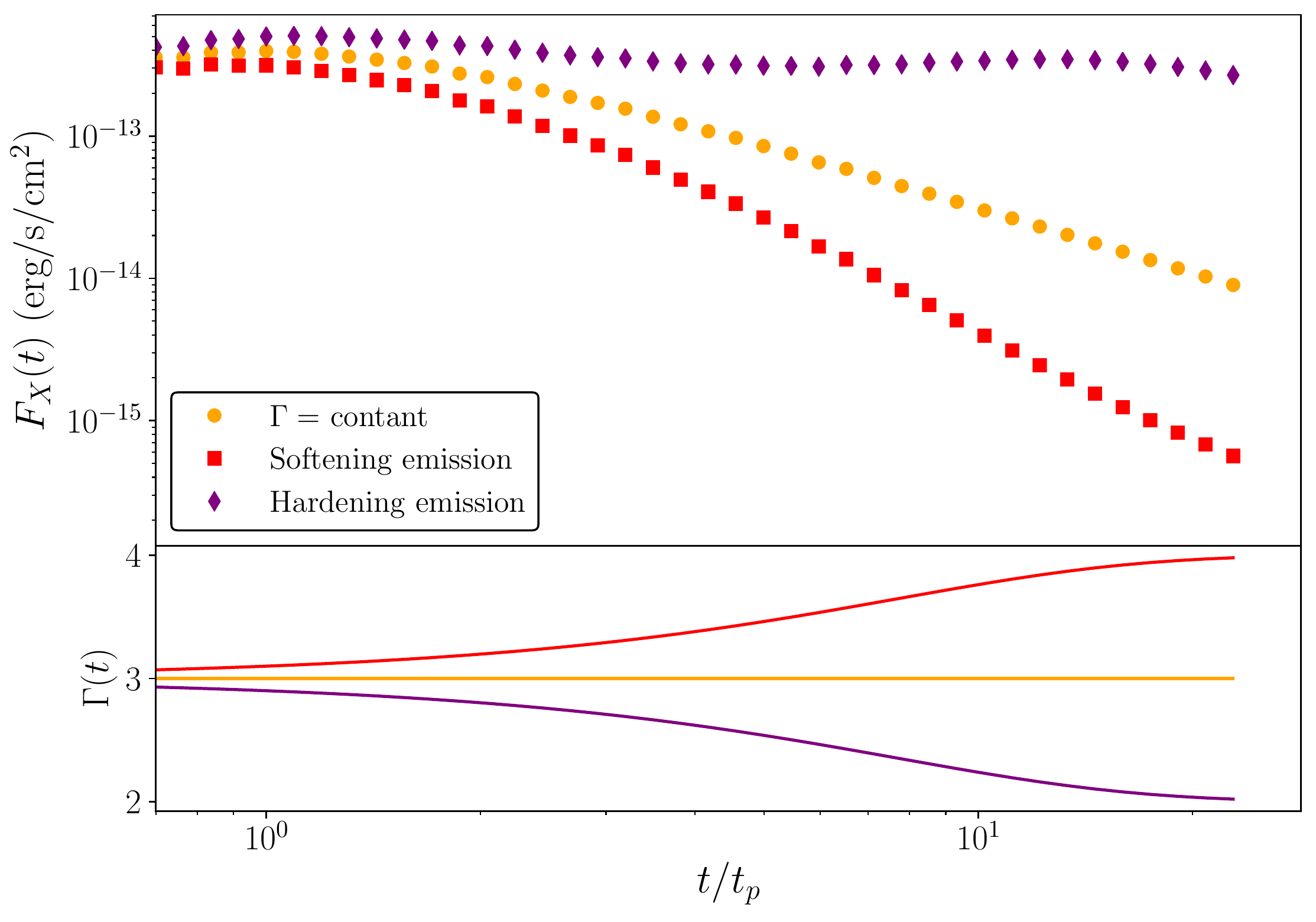} 
 \caption{The effect of time-dependent photon indices $\Gamma(t)$ on the evolving nonthermal  X-ray light curves of simple disc models. Corona with photon indices which harden (decreasing $\Gamma$, purple diamonds and curve) remain significantly brighter than constant $\Gamma$ models, while softening (increasing $\Gamma$, red squares and curve) photon index light curves  decay at a more rapid rate. All light curves were produced with $M_d = 0.5 M_\odot$, $\alpha = 0.1$ and $M = 2\times10^7 M_\odot$.  } 
 \label{change_lc}
\end{figure}

\begin{figure}
    \includegraphics[width=.5\textwidth]{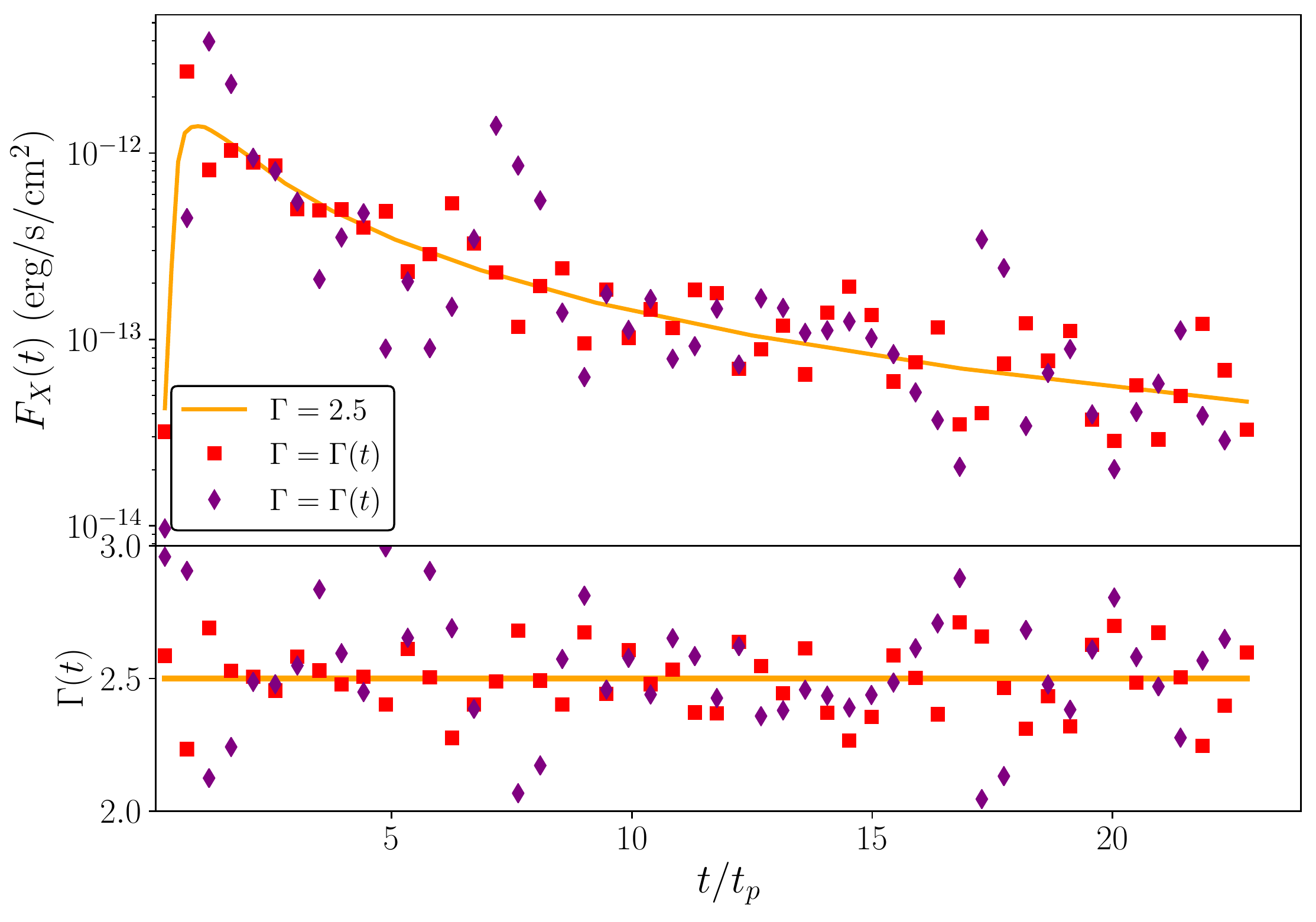} 
 \caption{The effect of time-dependent photon indices $\Gamma(t)$ on the evolving nonthermal  X-ray light curves of simple disc models.  Note that small time-scale fluctuations in $\Gamma$ of order $\sim 10\%$ cause much larger relative fluctuations in the observed X-ray flux (e.g., purple diamonds).   } 
 \label{rand_lc}
\end{figure}

The above light curves were produced numerically assuming that the properties of the corona were time-independent.   In reality, there is no powerful physical reason why this must be the case.   TDEs have been observed with nonthermal  X-ray spectral components which both soften (e.g., AT2019azh, Wevers 2020), and harden (e.g., AT2018fyk, Wevers 2020) at late times.  Significant evolution of the $\Gamma$ parameter can cause large deviations from the simple power-law light curves of Figure \ref{fid_lc}.   This is demonstrated in Figures \ref{change_lc} \& \ref{rand_lc}, which examine some simple ways in which the coronal properties may vary.

Figure \ref{change_lc} shows three 0.3 -- 10 keV light curves for accretion discs with an identical disc evolution profile.   Each disc has initial mass $M_d = 0.5 M_\odot$ and $\alpha = 0.1$, and surrounds a Schwarzschild black hole of mass $M = 2 \times 10^7 M_\odot$.  The only difference between the light curves stems from the evolution of the coronal $\Gamma$ parameter, which in this instance smoothly vary over the course of the disc evolution according to 
\beq
\Gamma(t) = \Gamma_0 \pm \delta \Gamma \tanh\left({t / 10t_p}\right) . 
\eeq
We take $\Gamma_0 = 3$, $\delta \Gamma = 1$, and $+/-$ corresponds to softening/hardening emission. The evolving $\Gamma$ profiles are shown in the lower half of Figure \ref{change_lc}.  As well as determining the rate at which the X-ray flux decays, the $\Gamma$ parameter also influences the amplitude of the X-ray flux itself.  This results in the softening (red) light curve falling off at a much more rapid rate, while the hardening (purple) light curve re-brightens at large times.

Furthermore, if the $\Gamma$ parameter varies on short timescales,  then much larger amplitude fluctuations in the observed X-ray light curves can be observed.   Figure \ref{rand_lc} shows the effect of random fluctuations in the $\Gamma$ parameter on the observed X-ray light curves. The disc evolution is identical to the light curves of Fig. \ref{change_lc},  but now the $\Gamma$ parameter is given by $\Gamma = 2.5$ (orange curve), or is normally distributed around $\Gamma = 2.5$ with standard deviation equal to $0.05\Gamma$ (red points) or $0.1 \Gamma$ (purple points). 
While the time-dependent $\Gamma$ profiles of Figs. \ref{change_lc} \& \ref{rand_lc} need not be a central feature of real TDE systems, it is important to note the senstivity to time-dependent coronal properties on the observed light curves of smoothly evolving disc systems.

\section{X-ray TDE Population analysis}\label{population}
\subsection{Overview}
Our analysis predicts that the populations of thermal X-ray TDEs and nonthermal  X-ray TDEs should be drawn from black hole populations with systematically different masses, distributed either above (nonthermal ) or below (thermal) a characteristic black hole mass scale of order $\sim 10^7 M_\odot$. 
In this section, we compare the currently known distribution of nonthermal  X-ray TDE masses with the predictions of this paper; namely, that the black hole mass distribution should peak above $M_p > 10^7 M_\odot$ and  decline at lower masses, with sources of mass $M \lesssim 5\times 10^6 M_\odot$ being rare.    The black hole mass $M = 5\times 10^6 M_\odot$ corresponds to the hard state mass scale for a disc with small initial mass content $M_d = 0.05 M_\odot$, and $\alpha = 0.1$.  

The authors are aware of seven X-ray bright TDEs with X-ray spectra which are well-modelled by nonthermal emission, and which have formal estimates of the central black hole mass. For this analysis we use well-established 
galactic scaling relationships between the black hole mass and (i) the galactic bulge mass $M : M_{\rm bulge}$, (ii) the galactic velocity dispersion $M : \sigma$, and (iii) the bulge V-band luminosity $M : L_V$. All of the scaling relationships are taken from McConnell \& Ma (2013).  Where available, values of $ M_{\rm bulge}$, $\sigma$ and $L_V$ were taken from the literature for each TDE, and are presented in Table \ref{fulldatatable} in Appendix \ref{data}. The mean black hole mass for each TDE is given in Table \ref{datatable}.  

{
\subsection{Source selection and hard state designation} 

\subsubsection{Source selection} 
Our sample selection criteria is very simple.   For inclusion, the TDE must have two properties: an unambiguous early-time X-ray detection with a well-observed X-ray spectrum.   In addition, there must at least one published estimate for the central black hole mass.   Note that we include only those TDEs which are X-ray bright at early times, and neglect those that were initially X-ray dim but then observed as X-ray bright only at much later times.   (There are three such sources reported in Jonker {\it et al}. 2020).  The reason for this neglect is that the focus here is upon those TDEs which have formed around large mass black holes at low Eddington ratios, rather those which have {\it transitioned} into the hard state as a result of cumulative late-time cooling.

\subsubsection{Hard and soft state designation} 
The decision of whether a TDE should be designated as a hard state TDE or a soft state TDE is based, of course, entirely upon the properties of the sources X-ray spectrum. For many sources this was unambiguous, as they were observed in either a `pure' thermal or nonthermal state.  However, other sources have both thermal and nonthermal components present in their X-ray spectrum, and some additional consideration is required.   The X-ray spectra of all of these `mixed sources'  was quantitatively analysed as part of the initial data reduction, and the fraction of the total observed X-ray flux resulting from each spectral component determined (e.g. Table 1, Wevers 2020). 
In mixed cases,  if the X-ray flux is more than $50\%$ dominated by power-law emission,  it is designated as `hard state'; otherwise it is designated as `soft state'. For sources where there were multiple observations with spectral information present we use the observation at which the source is at its maximum X-ray luminosity to designate its spectral classification. 

This procedure sorts 18 of the 19 TDEs with both black hole mass estimates and a well-observed X-ray spectrum.   There is one potentially ambiguous source of interest, AT2018fyk, which could arguably be sorted into either category,  and is discussed below.  
 
\subsubsection{Edge cases: AT2018fyk and XMMSL1 J0740} 
Only one source has been observed to transition between accretion states at early times,  AT2018fyk (Wevers {\it et al}. 2019).   AT2018fyk was initially observed with a two-component X-ray spectrum with  $\sim 25 \pm 13\%$ of its X-ray flux resulting from its power-law component, suggesting that it is a soft-state TDE.  However, AT2018fyk then displayed behaviour unlike any other soft state TDE, and rapidly transitioned into the hard state, with a power-law fraction of $\sim 80\%$ (Wevers {\it et al}. 2021). AT2018fyk transitioned into a power-law dominated state just 75 days after it was initially detected (Wevers {\it et al}. 2021).  {Of course, state transitions within accretion discs are not expected to be instantaneous, indeed they may occur over several thermal timescales. When exactly AT2018fyk left one state and entered another is therefore hard to pin down.  } 

In this work we have opted to designate AT2018fyk a hard-state TDE, for two primary reasons. First, there is hard nonthermal emission present at all times during its evolution, suggesting that a corona formed in the earliest stages of the disc evolution. Second, AT2018fyk is at maximal X-ray brightest when it has transitioned fully into its corona-dominated state.  

What is clear is that AT2018fyk must have formed with a small Eddington ratio, very close to the transitional luminosity scale ($\sim 0.01 L_{\rm edd}$).  The case for this is fairly compelling, as the viscous evolution timescale of an accretion disc with black hole mass $M \simeq 4\times10^7 M_\odot$ is of the order $\gtrsim 100$'s of days.  For AT2018fyk to have completed its state transition in less than 100 days, it must have begun its evolution (at the very least) already close to the state-transition luminosity scale. It is therefore of some interest that AT2018fyk is the most massive black hole of those which formed with $< 50\%$ X-ray flux coming from the power-law when first observed, precisely as would be expected if initial Eddington ratio is strongly (inversely) correlated with central black hole mass.  

Finally, the source XMMSL1 J0740 is also a potential borderline case, with an observed power-law fraction of $f_{PL} = 59 \pm 11 \%$ (Wevers 2020), meaning that, within the quoted uncertainty, it could be deemed a soft state TDE.   It is perhaps unsurprising, therefore, that XMMSL1 J0740 has the smallest inferred black hole mass of all of the hard-state TDEs in the literature. This small black hole mass is in keeping with what would be expected from the analysis performed in this paper.  

\subsubsection{Possible contamination of TDE sample from flaring AGN}
Three of the sources included in the `hard state' TDE sample (XMMSL1 J0619, Saxton {\it et al}. 2014; ASASSN-18jd, Neustadt {\it et al}. 2020; and PTF-10iya, Cenko {\it et al}. 2012), have also been flagged as potential flaring AGN by the discovering authors.   The difficulty here is that, unlike the super-soft X-ray flares from soft-state TDEs which are easily distinguished from variable AGN behaviour, TDEs which form discs in the hard state produce X-ray spectra much more similar to those of AGN {(see Zabludoff {\it et al}. 2021 for further discussion)}.  It is therefore a challenge to distinguish between a TDE which forms a disc in a harder accretion state, and a flare which takes a previously unobservable (but nevertheless present!) AGN disc to a higher, and thereafter observable, luminosity. 

Distinguishing between flaring AGN and TDEs which form in a hard accretion states will always be a difficult observational problem, one that will not be solved in this work.    The reader should be aware, therefore, that not every classification in the samples is completely secure, and that our conclusions must be interpreted bearing in mind this caveat.  
}
\subsection{Results}
{Noting the above uncertainties, we next analyse the two X-ray sub-populations of TDEs.   In Table \ref{datatable} we present the mean black hole masses of the 7 TDEs with dominant nonthermal (hard state) emission, in Table \ref{datatabletherm} we present the mean black hole masses of the 12 TDEs dominated by thermal (soft state) emission.

}

\begin{table}
\renewcommand{\arraystretch}{2}
\centering
\begin{tabular}{|p{2.2cm}|p{2.cm}|}\hline
TDE name  & $\left\langle M_{\rm BH}\right\rangle/10^6M_\odot$  \\ \hline\hline
ASASSN-18jd & $ 145^{+105}_{-116} $  \\ \hline
AT2018fyk & $ 37.7^{+53.4}_{-12.0} $ \\ \hline
XMMSL1 J0740 & $7.9^{+4.4}_{-2.9}  $ \\ \hline
XMMSL2 J1446 & $41.2^{+47.1}_{-27.9}  $\\ \hline
SDSS J1323 & $14.7^{+20.7}_{-10.2}  $ \\ \hline
PTF-10iya & $34.9^{+46.7}_{-22.6}  $ \\ \hline
XMMSL1 J0619 &$33.3^{+19.6}_{-19.6}  $ \\ \hline
\end{tabular}
\caption{The  mean black hole mass of the 7 nonthermal (hard state)  X-ray TDEs from the literature.  See Appendix \ref{data} for further details. }
\label{datatable}
\end{table}

\begin{table}
\renewcommand{\arraystretch}{2}
\centering
\begin{tabular}{|p{2.2cm}|p{2.cm}|p{2cm} |}\hline
TDE name  & $\left\langle M_{\rm BH}\right\rangle/10^6M_\odot$ \\ \hline\hline
ASASSN-14li & $2.9^{+2.9}_{-1.6} $  \\ \hline
ASASSN-15oi & $8.1^{+7.1}_{-4.3}$ \\ \hline
AT2018hyz & $4.3^{+6.9}_{-3.3} $ \\ \hline
AT2019dsg & $20.4^{+28.3}_{-14.7} $ \\ \hline
AT2019azh & $4.5^{+8.0}_{-3.5} $ \\ \hline
AT2019ehz &$6.6^{+8.0}_{-3.8}$ \\ \hline
AT2018zr & $11.0^{+14}_{-6.7}  $ \\ \hline
SDSS J1311 & $5.2^{+8.9}_{-3.3} $ \\ \hline
XMMSL1 J1404 & $2.8^{+1.4}_{-1.0} $ \\ \hline
OGLE 16aaa & $26.0^{+35}_{-16} $ \\ \hline
3XMM J1521 & $5.4^{+5.1}_{-3.0}$ \\ \hline
3XMM J1500 & $7.3^{+5.5}_{-3.2}$ \\ \hline
\end{tabular}
\caption{{The mean black hole mass of the 12 Thermal (soft state) X-ray TDEs from the literature. This table is reproduced from Paper I. The mean mass values were calculated in an identical manner to the hard-state TDEs in Table \ref{datatable}. } }
\label{datatabletherm}
\end{table}

\begin{figure}
  \includegraphics[width=.5\textwidth]{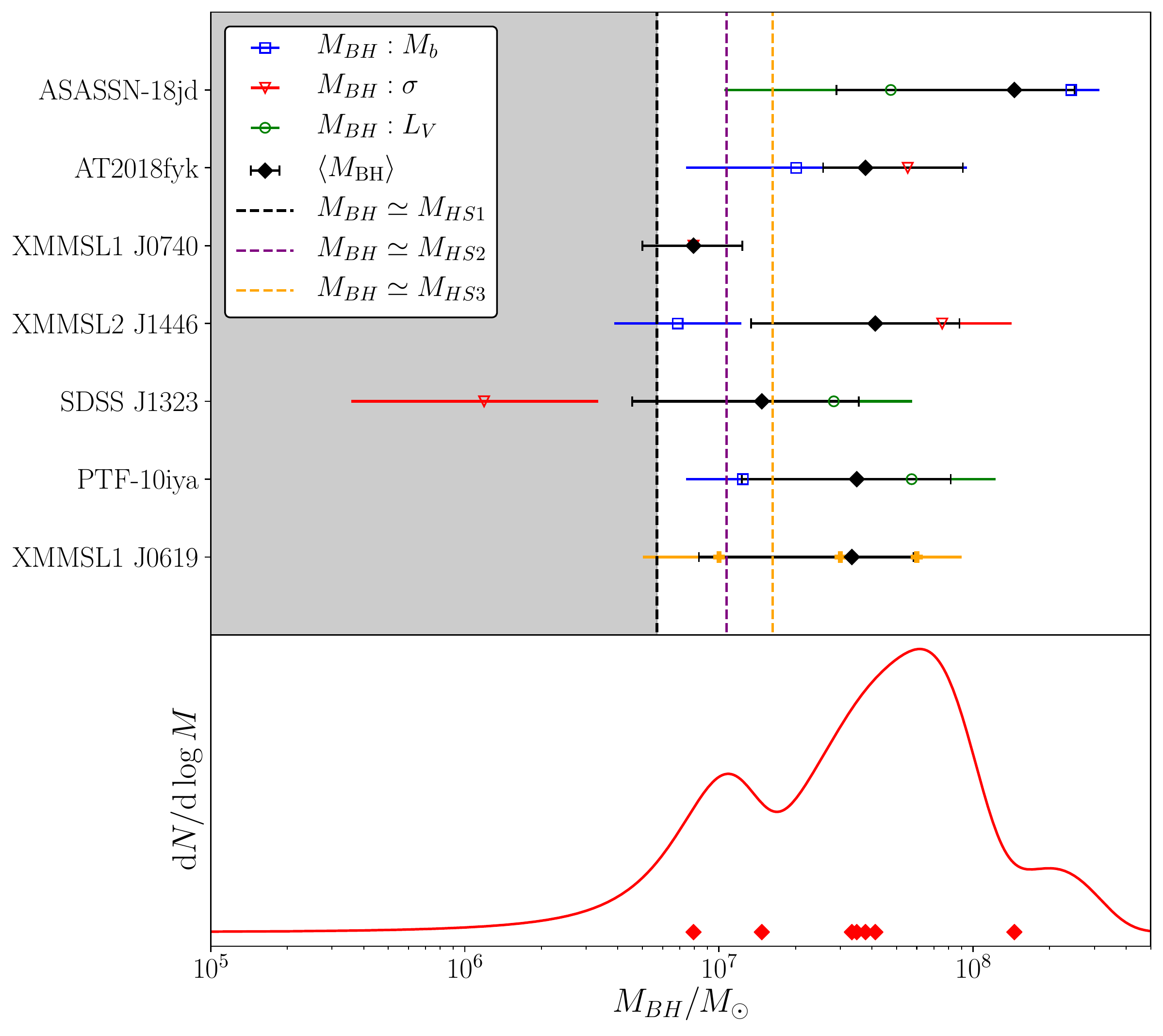} 
 \caption{The black hole mass distribution of TDEs with bright nonthermal  X-ray spectra. The upper panel shows the black hole mass values inferred for each individual TDE, calculated using correlations with  (i) the galactic bulge mass $M : M_{\rm bulge}$ (blue squares), (ii) the galactic velocity dispersion $M : \sigma$ (red triangles), and (iii) the bulge V-band luminosity $M : L_V$ (green circles). The mean black hole mass for each TDE is plotted as a black diamond.  The lower panel shows the black hole mass distribution of the nonthermal  X-ray TDE population, obtained using kernel density estimation using a kernel width equal to the uncertainty in each TDEs black hole mass. } 
 \label{nonthermpop}
\end{figure}

In the upper section of figure \ref{nonthermpop} we show the black hole masses of the seven nonthermal  X-ray TDEs obtained from each galactic scaling relationship, along with the mean black hole mass of each TDE (black diamond). We display as vertical dashed lines three characteristic hard state (Schwarzschild) black hole masses, corresponding to initial disc masses $M_d = 0.05, 0.2$ and $0.5 M_\odot$,  with $\alpha = 0.1$,  under the assumption that discs form in the hard state when the Eddington ratio is $l < 0.01$. 
In the lower section of figure \ref{nonthermpop} we show the current distribution of the black hole masses of the nonthermal  X-ray TDE population, obtained using kernel density estimation.   Explicitly, we use a Gaussian kernel with kernel width equal to the uncertainties in $M$ quoted in Table \ref{datatable} to represent each measurement as a probability density function (pdf). These pdfs are then summed over the relevant samples to obtain the distributions. 

The nonthermal  X-ray TDE population appears to be completely consistent with the expected results of this paper:  the distribution peaks at a black hole mass $M \simeq 6 \times 10^7 M_\odot$, and is strongly suppressed below $M \lesssim 1 \times 10^7M_\odot$.

\begin{figure}
  \includegraphics[width=.5\textwidth]{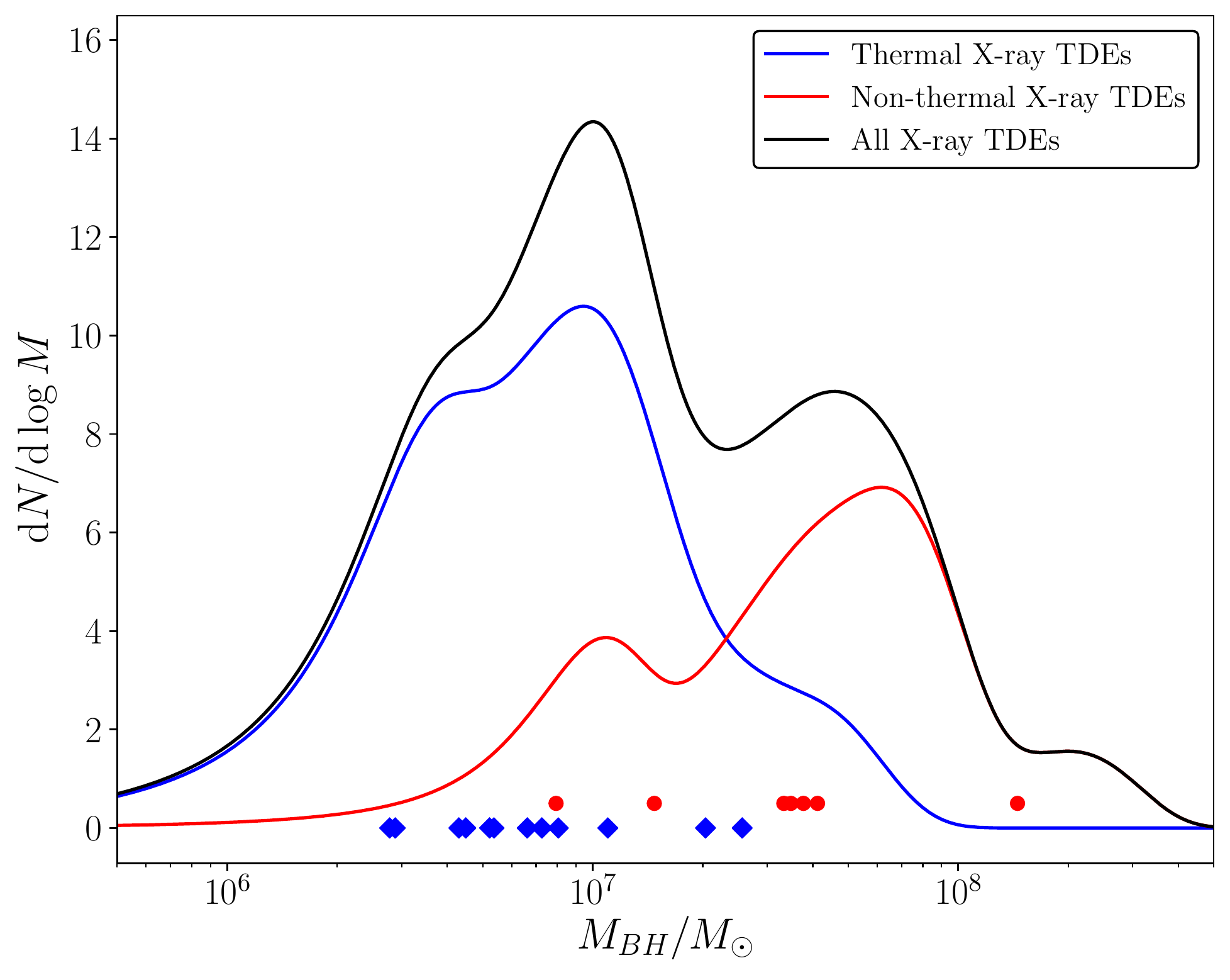} 
 \caption{The black hole mass distribution of the quasi-thermal (blue; Table \ref{datatabletherm}) \& nonthermal  (red; Table \ref{datatable}) X-ray TDE populations, obtained using kernel density estimation using a kernel width equal to the uncertainty in each TDEs black hole mass. The blue diamonds and red circles show the mean black hole mass values of the individual TDEs of the quasi-thermal and nonthermal  populations respectively.  There is a clear systematic mass offset between the two sub-populations of X-ray bright TDEs, as predicted in this work.   } 
 \label{popcomp}
\end{figure}

In figure \ref{popcomp} we compare the current distributions of the black hole masses of both  the thermal and nonthermal  X-ray TDE populations, obtained using kernel density estimation.  The  kernel width is equal to the formal uncertainty in each TDEs black hole mass.   The population of thermal X-ray TDEs comprises of 12 TDEs, whose properties we have discussed in Paper I. Their  mean black hole masses were calculated in the same way (from galactic scaling relationships) as the nonthermal  TDEs in this paper, and the mean values are reproduced in Table \ref{datatabletherm}.     

Figure \ref{popcomp} is revealing.  While the total (thermal \& nonthermal) X-ray bright TDE population is approximately flat over a range of black hole mass between $M \sim 2\times 10^6 - 8\times10^7 M_\odot$ (as was first found by Wevers {\it et al}. 2019), there is a clear systematic mass offset between the distributions of the thermal and nonthermal  sub-populations, just as predicted.   Quantitatively, a two-sample K-S test rejects the null-hypothesis that the two sub-populations result from the same black hole mass distribution ($p$-value = 0.01).  A two-sample Anderson-Darling test (a test which is more sensitive to the wings of the distribution) also rejects this null-hypothesis, but at an even stronger level ($p$-value = 0.002). 

{ While the results of this analysis are clearly supportive of the model put forward in this paper, we note that by either changing the accretion state designation of the borderline TDE AT2018fyk, or by excluding it from the analysis entirely,  a lower statistical significance is found for the difference between the two sub-populations of TDEs.   Explicitly, excluding AT2018fyk results in $p$-values of $p = 0.03$ (K-S 2 test) and $p=0.004$ (Anderson-Darling test).   Changing AT2018fyk's accretion state designation results in $p=0.11$ (K-S 2 test) and $p=0.01$ (Anderson-Darling test).    Additional complications may arise if the two TDE subpopulations are contaminated by flaring AGN.   While the effect of misdiagnosing any particular subset of TDEs can be modelled and calculated, this is not particularly informative.    

Perhaps the best way to view these results is as promising initial support for a prediction which will be much more rigorously tested over the coming years.    The eRosita telescope, for example, is expected to find hundreds to thousands of X-ray bright TDEs over the next 5 years (Khabibullin {\it et al}. 2014, {Jonker {\it et al}. 2020}).   With this vastly increased sample size, it will be possible to perform a much more definitive analysis of the properties of the two TDE sub-populations. 

In summary, the properties of the known TDE population are thus far supportive of two distinct sub-populations of X-ray bright TDEs with different spectral properties, separated by black hole mass.  Explicitly, an analysis of the mean black hole masses of soft-state and hard-state TDEs with an Anderson Darling two-sample test shows statistically significant ($p < 0.01$) differences between their mass distributions. 

}

\section{A model of state transitions in TDE discs}\label{rebirghten}

{ The primary focus of the paper has been on those TDE discs which form in {\it either} the soft or hard state.  However, the behaviour of TDE sources which transition between accretion states is also of great astrophysical interest, as in the case of AT2018fyk, whose X-ray spectrum has become significantly harder as it has evolved (Wevers {\it et al}. 2021). }   In addition, as noted in \S\ref{massstates}, there are now observations of nonthermal  X-ray emission at very late times ($4-9$ years after discovery) from TDE systems, which at early times were X-ray dim (Jonker {\it et al}., 2020).    Although the actual transition between accretion states was not observed {for these sources}, this is nevertheless strong circumstantial evidence for an accretion state transition occurring in TDE accretion discs.  Indeed,  this late-time transition is to be expected, since the late-time bolometric luminosity of an accretion disc in standard models decays as a  power law, $L \sim t^{-n}$.   Discs which formed with Eddington ratios above the hard state transitional scale $l_{\rm peak} > l_{\rm HS}$, should transition when $l(t) < l_{\rm HS}$.  

To better understand the behaviour of the X-ray light curves of TDE accretion discs which undergo a state transition, we model the state transition of the disc at some Eddington ratio $l_u$ as a corona with an expanding coronal radius, in the following simple way:
\beq\label{switchon}
R_{\rm Cor}(l) = 
\begin{cases}
R_I \quad ({\rm no} \, {\rm scattering}), \quad &l(t) > l_u, \\
R_I + (R_C - R_I) f(l), \quad & l_u > l(t) > l_l, \\
R_C, \quad &l_l > l(t) ,
\end{cases}
\eeq
where 
\beq
f(l) = {l_u - l(t)  \over l_u - l_l}, \quad  l_u > l(t) > l_l.
\eeq

\begin{figure}
    \includegraphics[width=.5\textwidth]{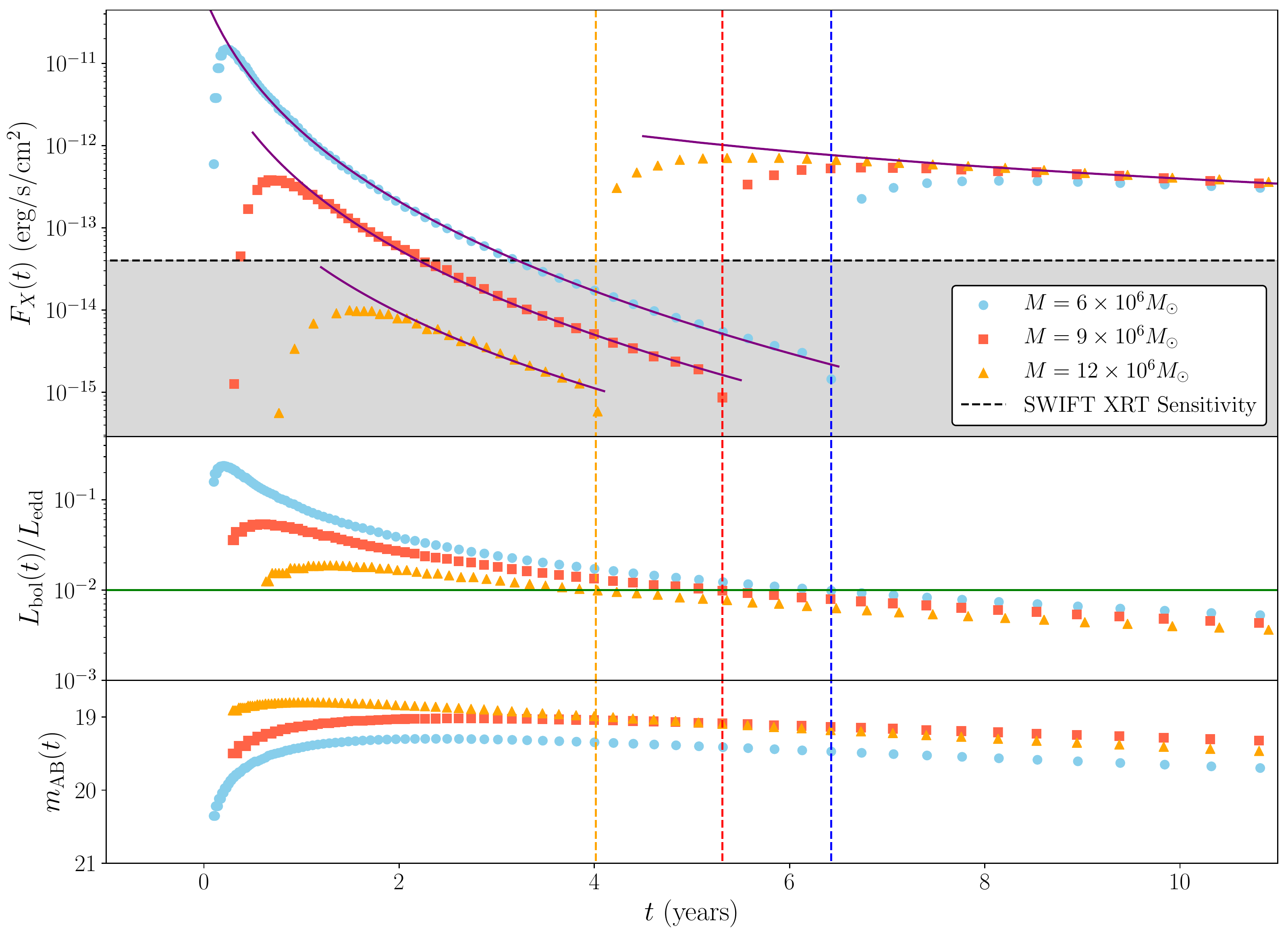} 
 \caption{Example X-ray (upper), bolometric (middle) and UV (lower) light curves from Schwarzschild black holes of different masses, denoted on plot. In this model the corona `switches on'  (equation \ref{switchon}) when the discs Eddington ratio becomes $l = 0.01$ (at times denoted by vertical dashed lines). The purple solid curves are the analytical models of equation \ref{lc} (late times, nonthermal  emission) and Mummery \& Balbus (2020a) (early times, thermal emission). Of particular interest is the orange curve, which is unobservable at X-ray energies at early times but becomes X-ray bright after $t > 4$ years.} 
 \label{switch_on_lc}
\end{figure}
Representative light curves are shown in Figure \ref{switch_on_lc}, in which three accretion discs, all with $M_d = 0.5 M_\odot$ and $\alpha = 0.1$, form around Schwarzschild black holes of different masses.   All have peak Eddington ratios $l_{\rm peak} > l_u = 0.01$.  The corona in each disc switches on when the evolving Eddington ratio of the disc $l(t_{u}) = l_u$, and then expands out to $R_{C} = 12r_g$ at $l(t_l) = l_l = 5\times10^{-3}$.   The corona scatters a fraction $f_{SC} = 0.3$ of the soft disc photons with a photon index $\Gamma = 3$.    

In the upper panel of Figure \ref{switch_on_lc} we show the X-ray light curves of the three disc solutions, in the middle panel the evolving Eddington ratios, and in the lower panel the disc UV light curves, expressed as an evolving AB magnitude for the UVW1 band. 
The discs are at a fiducial source distance $D = 100$ Mpc. The orange (triangular points) light curve is of particular interest, corresponding to the most massive black hole ($M = 12 \times 10^6 M_\odot$).   At early times ($t \lesssim 4$ years), the thermal disc emission is insufficient to produce observable levels of X-ray flux, due to the cool inner disc regions around such a large mass black hole (Paper I).  

However, at yet later times ($t \gtrsim 4$ years) the up-scattered nonthermal  X-ray emission is detectable.   {The properties of this light curve is therefore qualitatively similar to those observed by Jonker {\it et al}. (2020). We suggest that TDE sources which are initially X-ray dim and around black holes of large ($M \gtrsim 10^7 M_\odot$)  inferred masses may be interesting sources to follow up at X-ray frequencies at large times, as we expect them to transition to a harder accretion state within reasonable observational timescales. The detection of hard X-ray emission from these sources at large times will provide further evidence for the TDEs-as-scaled-up-XRBs paradigm. }

Finally, we note that the UV light curves, with their characteristic late-time plateau observed in many TDEs   (van Velzen {\it et al}. 2019, Mummery \& Balbus 2020a), are unaffected by the switching on of a compact corona (lower panel, Fig. \ref{switch_on_lc}).   This is because the UV emission is dominated by radii much farther out than the ISCO, and are thus unaffected by the inclusion of a compact corona in the inner disc regions. 

\begin{figure}
    \includegraphics[width=.5\textwidth]{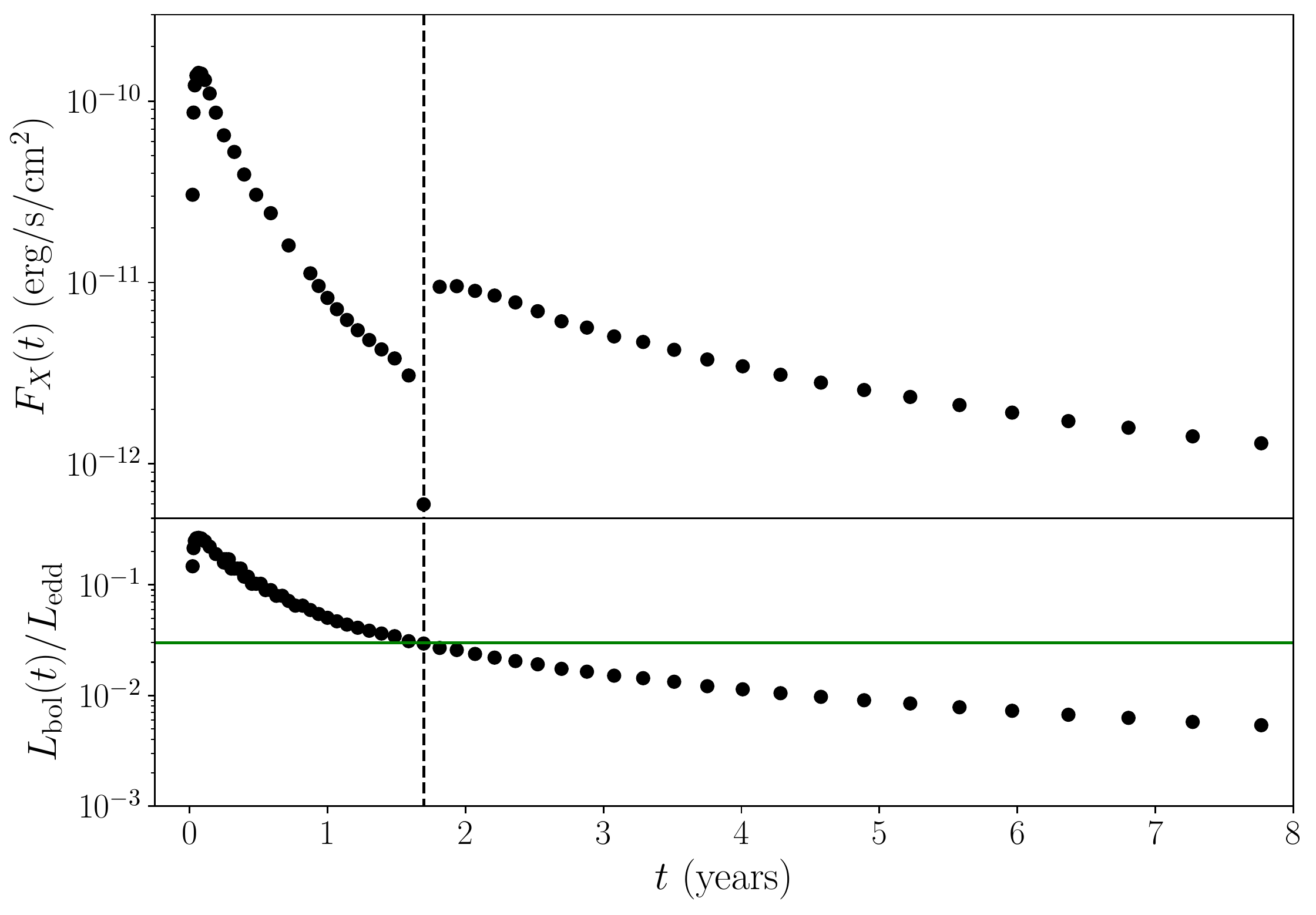} 
 \caption{The X-ray light curve of a rapidly rotating ($a/r_g = 0.9$) black hole of mass $M = 10^7M_\odot$. The disc parameters were $M_d = 0.1 M_\odot$ and $\alpha = 0.3$. The transition to harder emission occurred when the discs Eddington ratio was $l = 0.03$, at which point thermal X-ray emission was still observable. TDEs around black holes with large rotation parameters may offer the best opportunity to see TDE accretion disc transitions in real time.   } 
 \label{switch_on_lc_a09}
\end{figure}

{As highlighted in \S\ref{results}},  in contrast to the discs {of figure \ref{switch_on_lc}}, where the actual accretion state transition occurs at times when unobservable levels of X-ray flux are produced, discs around more rapidly rotating Kerr black holes will {be observed} to behave differently.   Figure \ref{switch_on_lc_a09} shows the X-ray and bolometric light-curves of a disc ($M_d = 0.1 M_\odot$, $\alpha = 0.3$), around a $M = 10^7 M_\odot$ black hole which is more rapidly rotating $a/r_g = 0.9$.  {These (physically reasonable) disc parameters were chosen so that the disc system underwent a state transition on the typical timescales of which TDE sources are observed.  As can be seen in figure \ref{switch_on_lc_a09},} this disc system begins to transition at $l(t_u) = 0.03$, when bright X-ray emission is observable. {It is interesting to note the very different rates of decay of the disc systems X-ray light curve in figure \ref{switch_on_lc_a09}, with the evolution much more rapid in the thermal-dominated state.  The existence of distinct rapid/slow decay modes is a general prediction of our model. } TDEs around more rapidly rotating Kerr black holes are the most promising systems to observe disc accretion state transitions in real time. 

\section{Key results}\label{conc}
In this paper we have extended the relativistic time-dependent thin-disc TDE model to describe nonthermal  X-ray emission produced by the Compton up-scattering of thermal disc photons by a compact electron corona. 

We have shown from simple scaling arguments that, assuming that TDE discs undergo state transitions into harder accretion states at Eddington ratios of order $L_{\rm bol} / L_{\rm edd} = l_{\rm HS} \sim 10^{-2}$,  nonthermal  X-ray emission will be detected from TDEs around black holes with masses larger than a characteristic scale $M_{\rm HS}$, which is given approximately by
\beq\label{MHS}
M_{\rm HS} \simeq 2\times10^7 \left({ M_d \over 0.5 M_\odot}\right)^{5/11} \left({ \alpha \over 0.1 }\right)^{4/11} \left({ l_{\rm HS} \over 10^{-2} }\right)^{-3/11} M_\odot ,
\eeq
where $M_d$ is the initial disc mass and $\alpha$ is the Shakura \& Sunyaev $\alpha$-parameter (1973).  A simple prediction of this analysis (in combination with our earlier work [Paper I]) is that nonthermal  and thermal X-ray TDEs should be drawn from different populations of black holes characterised by different masses, either being distributed above (nonthermal), or below (thermal), a characteristic black hole mass scale $M \sim 10^7 M_\odot$.  
We have demonstrated that the current X-ray bright TDE population (19 sources: 12 thermal, 7 nonthermal) are consistent with this picture. The mean black hole mass (calculated from galactic scaling relationships) of thermal TDEs is $\left\langle M_{\rm TH} \right\rangle = 9 \times 10^6 M_\odot $ while the mean black hole mass of nonthermal  X-ray TDEs is  $\left\langle M_{\rm NT} \right\rangle = 45 \times 10^6 M_\odot$.  A two-sample K-S test rejects the null-hypothesis that the two sub-populations result from the same black hole mass distribution ($p$-value = 0.01), while a two-sample Anderson-Darling test (a test which is more sensitive to the wings of the distribution) also rejects this null-hypothesis, but at an even stronger level ($p$-value = 0.002). 

We further demonstrated that the X-ray light curves of TDEs evolving in the hard state have the following large-time asymptotic behaviour 
\beq\label{anlc}
L_X(t) \propto t^{-(2 + \Gamma)n/4},
\eeq
where $\Gamma$ is the X-ray spectrum photon index, and $n$ is the bolometric luminosity decay exponent, $L_{\rm bol}(t) \sim t^{-n}$. 

Finally, we have developed a model where the electron corona `switches on' at large times for TDE discs which formed in the soft accretion state, but then transitions into a harder accretion state as matter is accreted and the disc luminosity decays with time. 

{This work represents the first detailed modelling of evolving TDE light curves which includes both thermal and nonthermal X-ray components.  By drawing connections between TDE discs and the better established properties of XRB discs, we have been able to make detailed predictions for the properties of the population of hard state X-ray TDEs.  These predictions can and will be tested rigorously in the near future, as considerably more observational data become available.  If confirmed, the work will have established a connection between TDE and XRB discs,  accreting onto black holes which span a range of 100 million in mass. }

\section*{Data accessibility statement}
All data used in this paper is presented in full in Appendix \ref{data}. 

\section*{Acknowledgments}
This work is partially supported by STFC grant ST/S000488/1, and the Hintze Family Charitable Foundation. It is a pleasure to acknowledge useful conversations with Adam Ingram, and extremely helpful comments from our referee.

\appendix{}
\section{TDE black hole masses from galactic scaling relationships}\label{data}

\begin{table*}
\renewcommand{\arraystretch}{2}
\centering
\begin{tabular}{|p{2.cm}|p{2.cm}| p{2. cm} | p{1.5 cm} | p{2. cm} | p{2. cm} | p{2. cm} | p{1.5 cm} |}
\hline
TDE name  & $M_{\rm bulge}/M_\odot$  &  $M_{\rm BH}/M_\odot$ & $\sigma$ (km/s)  &  $M_{\rm BH}/M_\odot$ & $L_{V}/L_\odot$  &  $M_{\rm BH}/M_\odot$ &Reference \\ \hline\hline
ASASSN-18jd & $8.5^{+0.05}_{-0.45} \times 10^{10}$ & $2.4^{+0.7}_{-1.6} \times 10^{8}$ & --- & --- & $4.0^{+4.0}_{-2.0} \times 10^9$ & $4.7^{+13.2}_{-3.7} \times 10^7$ & [1] \\  \hline
AT2018fyk & $7.9^{+17.2}_{-2.9} \times 10^{9}$ & $2.0^{+7.4}_{-1.3} \times 10^{7}$ & $158 \pm 1$ & $5.5^{+1.4}_{-1.1} \times10^7 $  & --- & --- & [2], [3] \\  \hline
XMMSL1 J0740 & --- & --- & $112\pm 3$ & $7.9^{+4.4}_{-2.9} \times10^6 $ & --- & --- & [4], [3]  \\  \hline
XMMSL2 J1446 & $2.9 \times 10^{9}$ & $ 6.9^{+5.3}_{-3.0} \times 10^{6}$ & $167 \pm 15$ & $7.6^{+6.6}_{-4.0} \times 10^7$ & --- & --- & [5], [3] \\  \hline
SDSS J1323 & --- & --- & $80 \pm 10$ & $1.2^{+2.2}_{-0.83} \times 10^6 $ & $2.5\times10^9$ &  $2.8^{+3.0}_{-1.4} \times 10^7$ & [6] \\  \hline
PTF-10iya & $5.0 \times 10^{9}$ & $1.2^{+0.9}_{-0.5} \times 10^{7}$ & --- & --- & $4.7^{+0.7}_{-0.6}\times10^9$ &  $5.7^{+6.6}_{-3.1} \times 10^7$ & [7] \\  \hline
\end{tabular}
\caption{The properties of the central black hole of  6 nonthermal  X-ray TDEs form the literature. [1] Nesutadt {\it et al}. (2020), [2] Wevers {\it et al}. (2019), [3] Wevers (2020), [4] Saxton {\it et al}. (2017), [5] Saxton {\it et al}. (2019), [6] Esquej {\it et al}. (2008), [7] Cenko {\it et al}. (2012).  }
\label{fulldatatable}
\end{table*}

\begin{table}
\renewcommand{\arraystretch}{2}
\centering
\begin{tabular}{|p{2.cm}|p{3.cm}| p{2. cm} | }
\hline
TDE name  & Measurement   &  $M_{\rm BH}/M_\odot$ \\ \hline\hline
XMMSL1 J0619 & K-band luminosity  & $1 {\pm 0.5} \times 10^7$ \\ \hline
XMMSL1 J0619 & H${}_\alpha$ luminosity  & $3 {\pm 1.5} \times 10^7$ \\ \hline
XMMSL1 J0619 & H${}_\beta$ line width  & $6 \pm 3 \times 10^7$ \\ \hline
\end{tabular}
\caption{The inferred black hole mass of the central black hole of XMMSL1 J0619 from three independent methods  (Saxton {\it et al}.\  2014). {Unfortunately no uncertainties were given on the K-band and H${}_\alpha$ measurements in (Saxton {\it et al}. 2014). In this work conservative $50\%$ error bars were chosen, which are equal to the quoted uncertainty for the H${}_\beta$ measurement.  } }
\label{XMM-J06}
\end{table}
To analyse  the current black hole mass distribution of thermal X-ray TDEs, we use well-established galactic scaling relationships between the black hole mass and (i) the galactic bulge mass $M : M_{\rm bulge}$, (ii) the galactic velocity dispersion $M:\sigma$, and (iii) the bulge V-band luminosity $M : L_V$.  
All of the scaling relationships are taken from McConnell \& Ma (2013).  Where available, values of $ M_{\rm bulge}$, $\sigma$ and $L_V$ were taken from the literature for each TDE, and are presented in Table \ref{fulldatatable}.

For some TDE hosts, only the total galactic mass was available,
 rather than the bulge mass. In these cases we assume a fixed fraction ($50\%$) of the host mass is in the bulge.   The uncertainty of each inferred black hole mass measurement is dominated by the intrinsic scatter of the galactic scaling relationships. 
Unfortunately, some host measurements have no reported uncertainties.   In these cases, the entirety of the black hole mass uncertainty results from intrinsic scatter in the scaling relationships. 

When a TDE has multiple independent black hole mass estimates we calculate the mean black hole mass via
$$
\left\langle M \right\rangle = {1\over N} \sum_{i=1}^N M_i.
$$
The (asymmetric) uncertainty on this  mean mass is taken to be
\beq
\left\langle \sigma^\pm \right\rangle = \left({1\over N}{\sum_{i=1}^{N} (\sigma^\pm_i)^2}\right)^{1/2} ,
\eeq
where $\sigma^+/\sigma^-$ correspond to the upper/lower uncertainties respectively. 
The mean black hole masses of each TDE candidate are displayed in Table \ref{datatable}.

The TDE XMMSL1 J0619 is an exception to the above analysis, as there were no reported values of $M_{\rm bulge}$, $\sigma$ or $L_V$ for this source. The reason we include this source in our analysis is that there exist three independent estimates of the central black hole mass for this TDE (Saxton {\it et al.} 2014), these estimates are reproduced in Table \ref{XMM-J06}.    With three independent estimates the inferred black hole mass of XMMSL1 J0619 should be reasonably accurate. 

\label{lastpage}

\end{document}